\documentclass[twocolumn]{aastex631}
\pdfoutput=1
\usepackage{amsmath,bm} 
\usepackage{chemformula}
\usepackage[utf8]{inputenc} 
\usepackage[T1]{fontenc}

\usepackage{afterpage}
\usepackage{xcolor}
\usepackage{afterpage}
\usepackage{mathtools}
\usepackage{xspace}
\usepackage{soul}
\usepackage{cancel}

\usepackage{booktabs}

\newcommand{\toie}{TOI-677\,b\xspace}
\newcommand{\toi}{TOI-3362\,b\xspace}
\newcommand{\water}{\ch{H2O}\xspace}
\newcommand{\methane}{\ch{CH4}\xspace}
\newcommand{\allplanets}{K2-139\,b, K2-329\,b, TOI-3362\,b, WASP-130\,b, WASP-106\,b, and TOI-677\,b\xspace}

\usepackage[separate-uncertainty = true,multi-part-units=single, group-digits = integer, group-four-digits = true,per-mode=reciprocal]{siunitx}

\newcommand{\numpm}[3]{$#1^{#2}_{#3}$}

\soulregister{\citep}{7}
\soulregister{\ch}{2}
\soulregister{\texttt}{1}
\soulregister{\water}{0}

\begin{document}

\title{High-resolution transmission spectroscopy of warm Jupiters: \\ An ESPRESSO sample with predictions for ANDES} 

\author{Bibiana Prinoth}
\affiliation{Lund Observatory, Division of Astrophysics, Department of Physics, Lund University, Box 118, 221 00 Lund, Sweden}
\affiliation{European Southern Observatory, Alonso de Córdova 3107, Vitacura, Región Metropolitana, Chile}

\author{Elyar Sedaghati}
\affiliation{European Southern Observatory, Alonso de Córdova 3107, Vitacura, Región Metropolitana, Chile}

\author{Julia V. Seidel}
\affiliation{European Southern Observatory, Alonso de Córdova 3107, Vitacura, Región Metropolitana, Chile}

\author{H. Jens Hoeijmakers}
\affiliation{Lund Observatory, Division of Astrophysics, Department of Physics, Lund University, Box 118, 221 00 Lund, Sweden}

\author{Rafael Brahm}
\affiliation{Facultad de Ingeniería y Ciencias, Universidad Adolfo Ibáñez, Av. Diagonal las Torres 2640, Peñalolén, Santiago, Chile}
\affiliation{Millennium Institute for Astrophysics, Chile}
\affiliation{Data Observatory Foundation}

\author{Brian Thorsbro}
\affiliation{Observatoire de la C\^ote d'Azur, CNRS UMR 7293, BP4229, Laboratoire Lagrange, F-06304 Nice Cedex 4, France}
\affiliation{Lund Observatory, Division of Astrophysics, Department of Physics, Lund University, Box 118, 221 00 Lund, Sweden}

\author{Andrés Jordán}
\affiliation{Facultad de Ingeniería y Ciencias, Universidad Adolfo Ibáñez, Av. Diagonal las Torres 2640, Peñalolén, Santiago, Chile}
\affiliation{Millennium Institute for Astrophysics, Chile}

\correspondingauthor{Bibiana Prinoth}
\email{bibiana.prinoth@fysik.lu.se}

\begin{abstract}

Warm Jupiters are ideal laboratories for testing the limitations of current tools for atmospheric studies. The cross-correlation technique is a commonly used method to investigate the atmospheres of close-in planets, leveraging their large orbital velocities to separate the spectrum of the planet from that of the star. Warm Jupiter atmospheres predominantly consist of molecular species, notably water, methane and carbon monoxide, often accompanied by clouds and hazes muting their atmospheric features. In this study, we investigate the atmospheres of six warm Jupiters \allplanets to search for water absorption using the ESPRESSO spectrograph, reporting non-detections for all targets. These non-detections are partially attributed to planets having in-transit radial velocity changes that are typically too small ($\lesssim 15$ km\,s$^{-1}$) to distinguish between the different components (star, planet, Rossiter-McLaughlin effect and telluric contamination), as well as the relatively weak planetary absorption lines as compared to the S/N of the spectra. We simulate observations for the upcoming high-resolution spectrograph ANDES at the Extremely Large Telescope for the two favourable planets on eccentric orbits, \toi and \toie, searching for water, carbon monoxide, and methane. We predict a significant detection of water and CO, if ANDES indeed covers the K-band, in the atmospheres of \toie and a tentative detection of water in the atmosphere of \toi. This suggests that planets on highly eccentric orbits with favourable orbital configurations present a unique opportunity to access cooler atmospheres.
\end{abstract}


\section{Introduction}
\label{sec:Intro}

\begin{figure*}[ht!]
    \includegraphics[width=\linewidth]{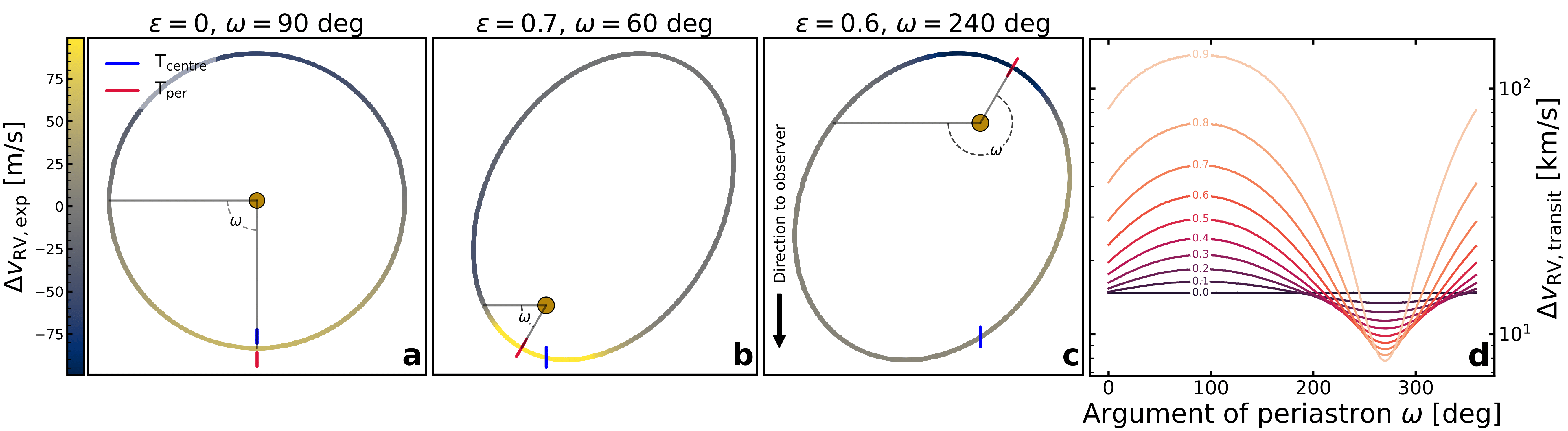}
    \caption{Change of planetary radial velocity for different orbital configurations per one-minute exposure (panels a-c) and from ingress to egress for varying eccentricities $\epsilon$ and argument of periastron $\omega$ (panel d). The planetary and stellar parameters are those of the TOI-3362 system, Table\,\ref{tab:planets}, but with varying eccentricity and omega as indicated.  \textit{Panel a:} Circular orbit ($\omega = 90 \deg$), resulting in a small radial velocity change during transit. \textit{Panel b:} Eccentric orbit with $\omega = 60 \deg$, leading to a large change in radial velocity during transit. \textit{Panel c:} Eccentric orbit with $\omega = 240 \deg$, resulting in a small change of radial velocity during transit. For the successful application of the cross-correlation technique, a large radial velocity change is desired to avoid overlap with the star's systemic velocity. \textit{Panel d:} Change in radial velocity during transit for the same system, with varying eccentricity $\epsilon$ and $\omega$. The radial velocity change is larger for highly eccentric orbits when the transit happens close or during periastron (e.g. panel b). The smallest change occurs when $\omega = 270\,\deg$, with the transit centre coinciding with apoastron. Below a certain eccentricity, it is preferable that the periastron passage does not coincide with the transit centre to distinguish components as the radial velocity goes through 0. 
    }
    \label{fig:graphics}
\end{figure*}

Warm Jupiters, Jupiter-like planets on orbits with periods longer than $\sim$\,10 days \citep{hinz_imaging_1998}, are optimal targets for pushing current methods of studying atmospheres to their limits. Unlike (ultra-)hot Jupiters, the atmospheres of these cooler siblings are expected to be predominantly composed of molecules such as water (\ch{H2O}), methane (\methane), carbon monoxide (\ch{CO}), and molecular nitrogen (\ch{N2}), see \citet{moses_chemical_2014}, and their atmospheric features are likely muted by clouds and hazes \citep[e.g.][]{lodders_exoplanet_2010, marley_clouds_2013,morley_quantitatively_2013}. Moreover, disequilibrium chemistry such as photochemistry \citep[e.g.][]{ahrer_early_2022,alderson_early_2022,feinstein_early_2022,rustamkulov_early_2022,tsai_global_2023} and transport-induced quenching \citep{moses_chemical_2014} become relevant, making these atmospheres challenging to study. 

A standard tool to search for atmospheric signatures of exoplanets is high-resolution transmission spectroscopy \citep{birkby_exoplanet_2018}. When the light emitted from the exoplanet's host star is filtered through the upper layers of the atmosphere of the planet, this leaves an imprint on the observed spectra, which later needs to be isolated to study the planet's atmospheric composition. While for hotter planets it is possible to observe them also in emission due to the presence of atmospheric inversion layers on the dayside \citep{fortney_hot_2021}, such layers are absent in planets with lower equilibrium temperatures \citep{garhart_statistical_2020}. Due to this lower temperature, and the inherent presence of clouds and hazes, transmission spectroscopy may be the only tool to directly access the atmospheric features of cooler planets, in particular warm Jupiters, as deep absorption lines may peak out above the cloud layer \citep{pino_diagnosing_2018}. One commonly used method to isolate planetary signatures is the cross-correlation technique that effectively sums up all lines of a given species. Originally introduced by \citet{snellen_orbital_2010} for studying exoplanet atmospheres, this technique is particularly effective for short orbits, as it uses the large radial velocity changes per unit time of the exoplanet relative to its host star. If the planetary orbital velocity is too small, indicating a distant orbit, the observed radial velocities of the planet and star may no longer be significantly separated during transit, potentially leading to contamination of the planetary trace by stellar components, or the Rossiter-McLaughlin (RM) effect, emerging from the planet covering the rotating stellar disk during transit \citep{rossiter_detection_1924,mclaughlin_results_1924}. Additionally, if the species of the planetary atmosphere are also present in the stellar photosphere, the planetary atmospheric signal would be located within the line cores of deep absorption lines, where there is little signal to measure its presence.

The study of cooler planet atmospheres dominated by molecular species that absorb strongly in the infrared wavelengths has gained focus thanks in part to advancements in current infrared instrumentation such as SPIRou \citep{donati_spirou_2020}, NIRPS \citep{bouchy_near-infrared_2017,wildi_nirps_2017} and CRIRES+ \citep{dorn_crires_2023}. This shift marks a departure from the intense recent studies of ultra-hot Jupiter atmospheres, which are exceptionally accessible due to their high temperatures and large radial velocities. Infrared observations are necessary to detect molecules in the spectra of cooler planets, which are characterised by longer orbital periods and smaller orbital velocities. Unlike their close-in counterparts, these more distant planets may not have undergone full circularisation of their orbits, due to the diminished influence of tidal forces exerted by the host star, which typically circularise the orbits of closer-in planets on relatively short timescales \citep{hut_tidal_1981}. Nevertheless, warm Jupiters are also observed to reside on (nearly) circular orbits as a result of in-situ formation, disk migration or circularisation timescales shorter than the age of the systems \citep[see][for a review]{dawson_origins_2018}.

One notable observational outcome of planetary formation processes is the emergence of warm Jupiter-like planets on highly eccentric orbits (e.g. TOI-4582b, HAT-P-17b, TOI-677b, Kepler-419b). These planets may have experienced violent interactions during their formation stages, and could still be undergoing high-eccentricity migration on the way to becoming hot Jupiters \citep[see][and references therein]{dawson_origins_2018}. Successful application of the cross-correlation technique to these eccentric systems depends on the ability to isolate planetary and stellar components, driven in particular through the eccentricity and the time of periastron, i.e. the time of the smallest distance between the planet and host star. While circular orbits are defined to have an argument of periastron of $90\,\deg$, in eccentric orbits they vary within the range of 0 to 360 $\deg$, depending on the system's orientation relative to an observer on Earth. For a highly eccentric orbit, if the timing of periastron passage is displaced relative to the transit centre, the planet's radial velocity is significantly shifted away from zero during transit, enabling the use of the cross-correlation technique to discern molecular species within its atmosphere, see Fig.\,\ref{fig:graphics} for a visual on the geometries. Conversely, if the periastron passage is near the superior conjunction, the radial velocity of the planet relative to the star during transit diminishes.


\subsection*{Observing exoplanets with the ELT}
With the construction of the Extremely Large Telescope (ELT) nearing completion, new instrumentation promises to revolutionise our ability to observe planetary atmospheres. ANDES (ArmazoNes high Dispersion Echelle Spectrograph) is a high-resolution spectrograph currently planned to be installed at the ELT as a second generation instrument \citep{Marconi2021,palle_ground-breaking_2023}. In many aspects, it will be the successor of ESPRESSO (Echelle SPectrograph for Rocky Exoplanets and Stable Spectroscopic Observation), the optical high-resolution spectrograph installed at the Very Large Telescope \citep{pepe_espressovlt_2021}\footnote{As well as other high resolution visible and near-infrared spectrographs, such as UVES, HARPS, NIRPS and (partially) CRIRES+}. ANDES will not only cover the wavelength range of ESPRESSO but will also extend further into the near-infrared regime, aiming to include K-band up to 2400\,nm. Together with the superior photon-collecting power of a 39m primary mirror (in contrast to 8.2m of a single unit telescope of the VLT), as well as being AO-assisted, the extended wavelength coverage will enable the detection of molecules like \water and \ch{CO}, as well as other species in the atmospheres of warm Jupiters on favourable orbits, significantly expanding the applicability of the cross-correlation technique to planets beyond orbital periods of 10\,days. Understanding the orbital configurations of these warm Jupiter-like planets will be crucial for allocating observational resources effectively, ensuring that only favourable configurations are prioritised for high-resolution cross-correlation studies.
\bigskip

\begin{figure}
    \centering
    \includegraphics[width=\linewidth]{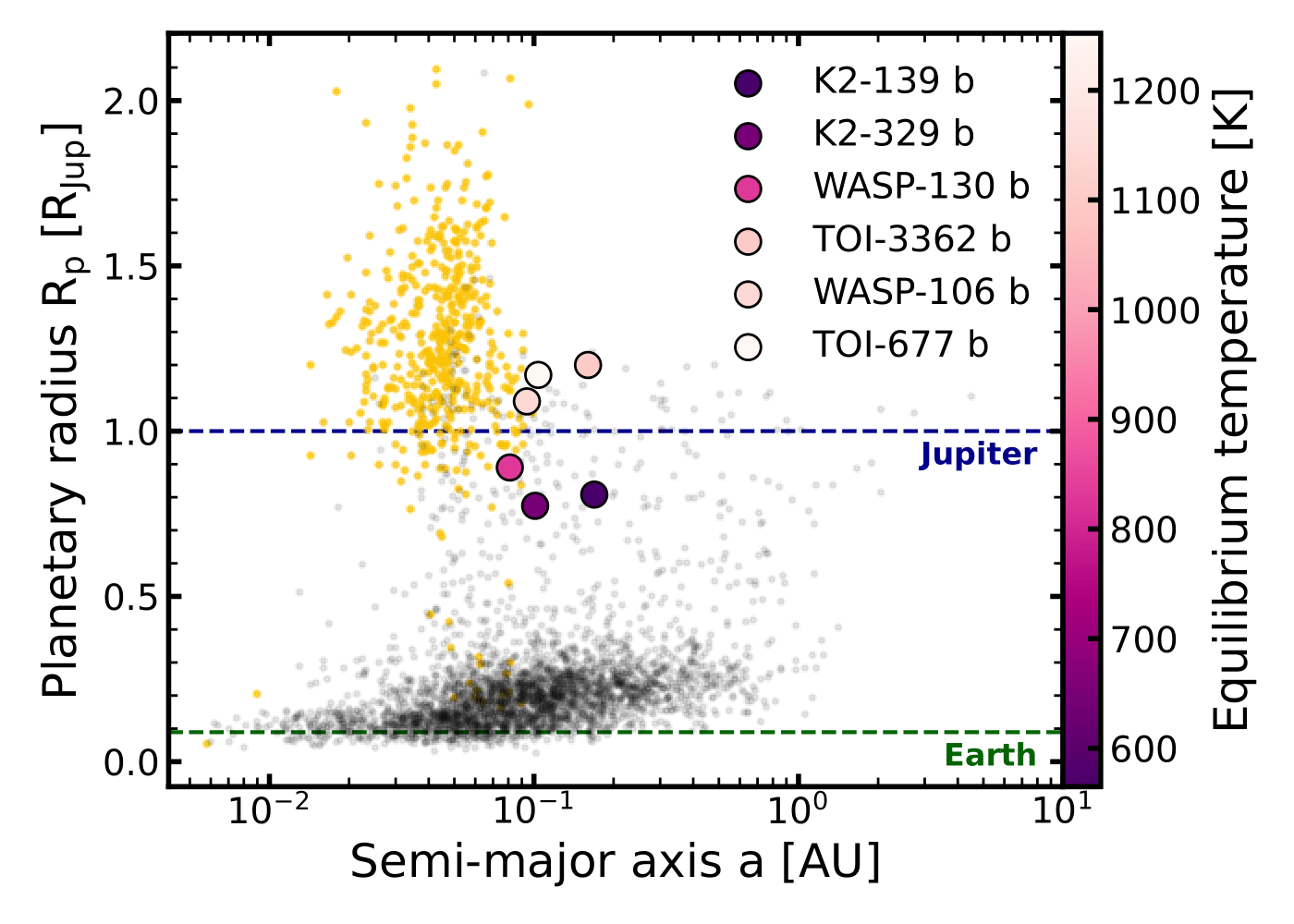}
    \caption{Planetary radius as a function of semi-major axis. The scatter points show the population of exoplanets discovered using transits for comparison, extracted from the NASA Exoplanet Archive on 19 February 2024. The yellow scatter points are considered hot and ultra-hot Jupiters  \citep [M$_{\rm p} \geq$ \SI{0.36}{M_{\rm Jup}}, P $\leq$ \SI{10}{days},][]{winn_transits_2010}. The radii of Jupiter and Earth are given as a reference.}
    \label{fig:intro_plot} 
\end{figure}

In this work, we present the analyses of the six warm Jupiters \allplanets, whose primary transits were observed with ESPRESSO. Our study focuses on searching for \water absorption at optical wavelengths in these datasets. We find that for some of these targets, either the orbital configuration is not favourable for detecting atmospheric absorption features due to the inability to distinguish planetary velocity from the stellar signal, or the signal-to-noise ratio (S/N) is insufficient to detect these atmospheres. For two planets, \toi and \toie, which are on eccentric orbits ($\epsilon \approx 0.7$ and $\epsilon \approx 0.4$, respectively, see Table \ref{tab:planets}), we explore the detectability of their atmospheres through model injection, leveraging their favourable orbital parameters. Additionally, we simulate observations for ANDES at the ELT, which could enable the detection of molecular species beyond \water, particularly \ch{CO}.

\begin{table*}[ht!]
    \caption{Overview of observations.}%
    \begin{center}
            \begin{tabular}{llllllll}
                    \toprule
                    Target      & Program ID / PI   & Date          & $\#$ Spectra$^a$  & t$_{\rm exp}$ [s] & Airmass $^{b}$ & Avg. S/N & Min./Max. seeing \\
                    \midrule
                    K2-139\,b   & 109.238M (Hobson) & 2022-07-05    & 23(18/5)          & 900               & 1.3-1.0-1.3   & 73.5      & 0.41 / 1.24 \\
                    K2-329\,b   & 109.238M (Hobson) & 2022-08-23    & 30(22/8)          & 600               & 1.1-1.1-1.9   & 41.6      & 0.44 / 0.78 \\
                    TOI-3362\,b & 110.23Y8 (Brahm)  & 2022-12-26    & 38(28/10)         & 310               & 1.7-1.2-1.2   & 38.6      & 0.88 / 1.39 \\
                    WASP-130\,b & 109.238M (Hobson) & 2022-06-08    & 47(28/19)         & 420               & 1.2-1.1-1.4   & 62.0      & 0.52 / 1.19 \\
                    WASP-106\,b & 110.23Y8 (Brahm)  & 2023-03-17    & 38(30/8)          & 600               & 1.3-1.1-2.2   & 56.7      & 0.51 / 1.08 \\
                    TOI-677\,b  & 108.22C0 (Brahm)  & 2021-12-09    & 62(41/21)         & 180               & 1.7-1.1-1.1   & 50.4      & 0.54 / 1.58 \\
                    \bottomrule
            \end{tabular}
    \end{center}
    \textit{Note:} $^{a}$ In parentheses, in-transit and out-of-transit spectra, respectively.  $^{b}$ Airmass at the start and end of the observation, as well as minimum airmass at the highest altitude of the target. 
    \label{tab:observation_log}
\end{table*}

The manuscript is structured as follows: In Section \ref{sec:obs}, we introduce the planetary sample and describe the observations, data reduction, and telluric correction. Section \ref{sec:methods} details our methodology, including the velocity corrections for elliptical orbits, the templates and models used for cross-correlation analysis and the approach to model injection for studying limitations with ESPRESSO, as well as the potential for observations with ANDES. In Section \ref{sec:results}, we present our findings and discuss their implications. Finally, in Section \ref{sec:Conclusions}, we conclude our study by summarising our main findings and discussing their broader significance for exoplanet atmospheres, particularly in light of advancements in observational capabilities.


\section{Observations \& data reduction}
\label{sec:obs}

The sample of planets in this study comprises of the six warm Jupiters \allplanets, orbiting F, G, and K stars at a distance of $\approx\SI{0.1}{AU}$. Their equilibrium temperatures range from \SI{565}{\kelvin} for K2-139\,b \citep[coldest planet in the sample,][]{barragan_k2-139_2018} to \SI{1252}{\kelvin} for TOI-677\,b \citep[hottest planet,][]{jordan_toi_677_2020}, see Fig.\,\ref{fig:intro_plot} for comparison to the transiting exoplanet population.

K2-139\,b, a low-density warm Jupiter, 
orbits an active K0\,V star in $\sim$\,29 days on a nearly circular orbit \citep[$\epsilon \approx 0.12$][]{barragan_k2-139_2018}. Its density is consistent with a solid core of \SI{49}{M_{\oplus}} from evolutionary models of \citet{fortney_planetary_2007}. 

K2-329\,b orbits a G dwarf star on a circular orbit every $\sim$\,12\,days \citep{sha_toi-954_2021}. The circularisation timescale for the system is of the order of the age of the universe, assuming its interior composition resembles that of Saturn \citep{lainey_new_2017}. However, the tidal quality factor depends strongly on interior structure and composition, so the tidal circularisation timescale could be an order of magnitude shorter, and be comparable to the age of the system. Alternatively, the circular orbit might be explained without tidal dissipation, which could rule out high-eccentricity migration, pointing instead to scenarios such as in-situ formation or disk migration \citep{sha_toi-954_2021}.

\toi is a proto-hot Jupiter that was found to reside on a highly eccentric orbit around a G-type star \citep[$\epsilon \approx 0.7$,][]{espinoza-retamal_aligned_2023}. Its orbital configuration suggests ongoing high-eccentricity migration, thought to lead to the eventual circularisation of its orbit closer to the host star as a hot Jupiter. This planet provides an intriguing opportunity to study the evolutionary path of hot Jupiter formation. According to \citet{espinoza-retamal_aligned_2023}, this observed orbital configuration may be attributed to co-planar high-eccentricity migration \citep{Petrovich2015}, influenced by the gravitational pull of a distant companion beyond \SI{5}{AU}.

WASP-130\,b orbits a metal-rich G6 star on a circular orbit with a period of $\sim$\,12 days \citep{hellier_wasp-south_2017}. Using VLT/SPHERE, \citet{bohn_multiplicity_2020} detected a companion candidate in the same system at a separation of 0.6 arc-seconds, which corresponds to a semi-major axis of approximately \SI{100}{AU}. If the companion is gravitationally bound, its mass is \numpm{0.3}{+0.3}{-0.2}\,$\si{M_\odot}$, making it likely a brown-dwarf companion.

WASP-106\,b is a warm Jupiter orbiting an F9 star on a circular and aligned orbit roughly every 9\,days \citep{stassun_accurate_2017,harre_orbit_2023,wright_soles_2023}. The circulation time-scale of this planet, estimated to be of the order of the system's age \citep{stassun_accurate_2017}, suggests that WASP-106\,b's orbit has not undergone circularisation from a highly eccentric starting point, but instead, it has likely maintained a nearly circular orbit throughout the system's lifetime. This combined with the aligned orbit hints at a quiescent formation pathway through disk migration \citep{harre_orbit_2023,wright_soles_2023}, i.e. type II. 

\toie resides on an eccentric ($\epsilon \approx 0.4$) 11-day orbit around its F-type host star, and shares similarities with \toi in terms of its eccentricity. Intriguingly, it also aligns with the projected spin-axis of its star, with \citet{sedaghati_orbital_2023} providing evidence suggesting that a far-out companion, a brown dwarf, likely does not influence its orbital configuration. \\

\bigskip
For all the host stars in this study, we conducted a homogeneous characterisation following a two-step iterative process initially presented in \citet{brahm:2019}. Firstly, we computed the stellar atmospheric parameters ($T_{\text{eff}}$, $\log{g}$, [Fe/H], and $v\sin{i}$) using the \texttt{zaspe} code \citep{zaspe}.  This involved comparing the co-added ESPRESSO spectra to synthetic ones in spectral regions most sensitive to changes in atmospheric parameters. Subsequently, we employed these parameters as priors in a spectral energy distribution (SED) fitting procedure. This procedure utilised public broadband photometry of each star, GAIA DR2 parallax, and PARSEC isochrones \citep{parsec}. We explored the parameter space using the \texttt{emcee} package to obtain posterior distributions for the stellar mass, radius, age, and interstellar extinction. These parameters enabled us to compute a more precise value for $\log{g}$, which was then used in a new run of \texttt{zaspe}. In this run, the $\log{g}$ parameter was held fixed to the value determined from the SED fit. We iterated this process until achieving convergence in $\log{g}$. All stellar parameters obtained through this procedure are summarised in Table\,\ref{tab:planets}, along with other relevant orbital and physical planet parameters obtained from the literature.


\bigskip

All our targets are consistent with the pattern that warm Jupiters in single-star systems show low spin-orbit alignment angles initially found by \citet{rice_tendency_2022}. However, further observations of warm Jupiter systems are necessary to draw a statistically robust conclusion.

For each target in the sample, a single primary transit was observed with ESPRESSO, where the primary fibre A was placed on the target, with Fibre B on sky. A more detailed log of the observations is provided in Table\,\ref{tab:observation_log}. The spectra were reduced using the dedicated data reduction pipeline (v.3.0.0), provided by ESO and the ESPRESSO consortium. The pipeline recipes which include bias and dark subtraction, flat-field correction, order definition, blaze and flux correction, sky subtraction, were run on the {\tt esoreflex} environment, provided by ESO. Subsequently, the pipeline provides a set of reduced products, including two-dimensional (order by order) spectra, both blaze and non-blaze corrected for fibre A and B, stitched and resampled one-dimensional spectra again for both fibres, flux calibrated fibre A one-dimensional spectrum, as well as the order by order cross-correlation functions calculated for each fibre. All spectra are provided including the dispersion solution, with wavelengths given both in air and vacuum, where the solution is determined using a combination of the Th-Ar and Fabry-Pérot frames taken as daytime calibrations.

\section{Methodology}
\label{sec:methods}

\begin{figure*}
    \centering
    \includegraphics[width=\linewidth]{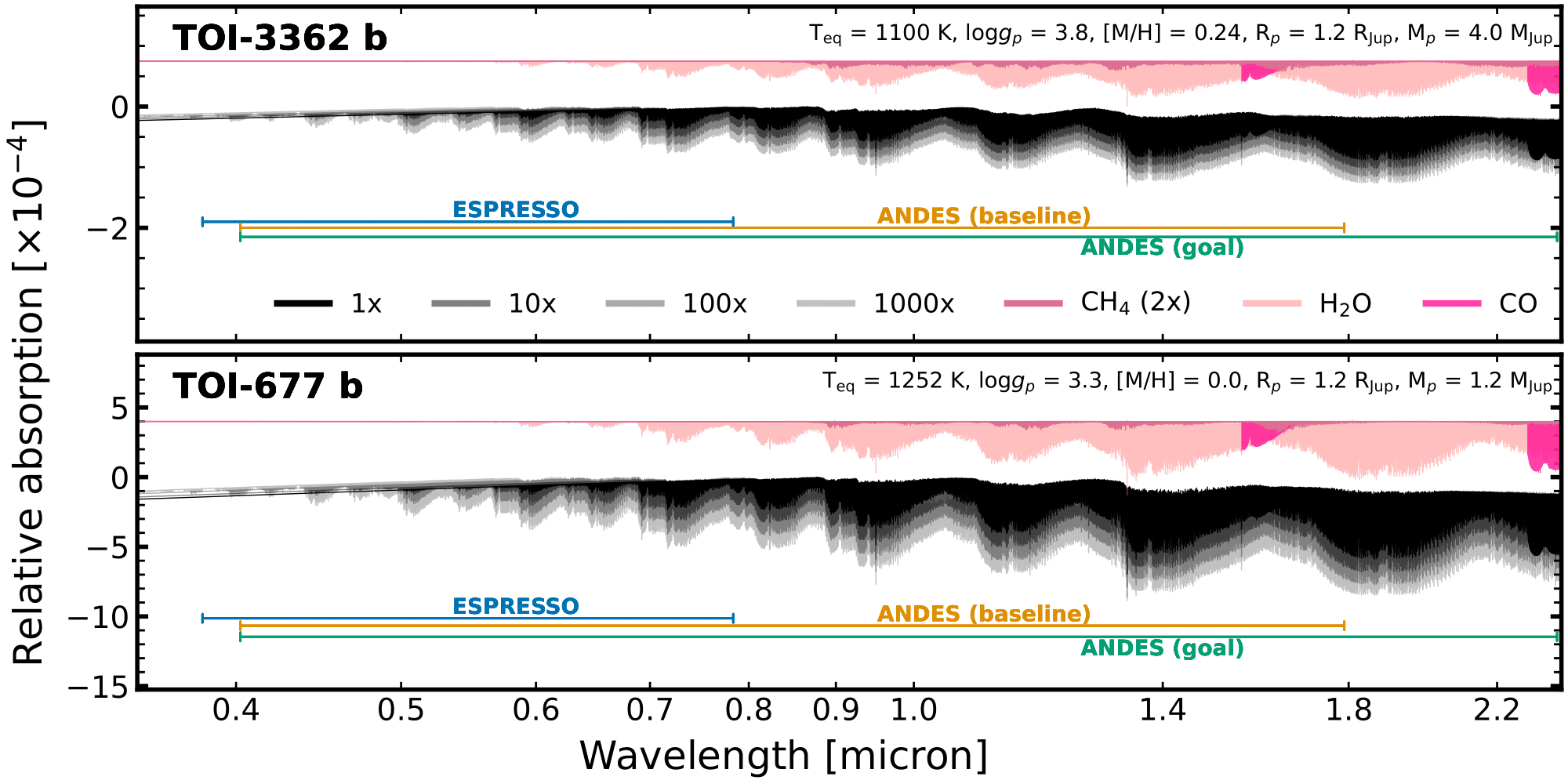}
    \caption{Transmission spectra models including \water, \methane and \ch{CO} of \toi (top) and \toie (bottom) in black with corresponding absorption bands of \water, \methane and \ch{CO} for reference in shades of pink. \textit{Top panel:} Model spectra of \toi including absorption of \water, \methane and \ch{CO}. The models assume an isothermal atmosphere at the calculated equilibrium temperature of $T_{\rm eq, A=0} = \SI{1100}{\kelvin}$. The corresponding abundance profiles were computed with \texttt{FastChem Cond} \citep{stock_fastchem_2018,stock_fastchem_2022,kitzmann_fastchem_2023}, see Fig.\,\ref{fig:fc_rainout}, and then translated into mass fractions to produce the transmission models using \texttt{petitRADTRANS} \citep{molliere_petitradtrans_2019}. The \ch{H2O} mass fractions were scaled with factors of 1 (nominal), 10, 100, and 1000 for model injection. Considered absorption bands included in the model to identify the features. The absorption band models were computed under the same assumption, but only including the considered species. \textit{Bottom panel:} Same as  top panel, but for \toie. The models assume an isothermal atmosphere at the calculated equilibrium temperature of $T_{\rm eq} = \SI{1252}{\kelvin}$ \citep{jordan_toi_677_2020}.}
    \label{fig:models}
\end{figure*}

\begin{figure}
    \centering
    \includegraphics[width=\linewidth]{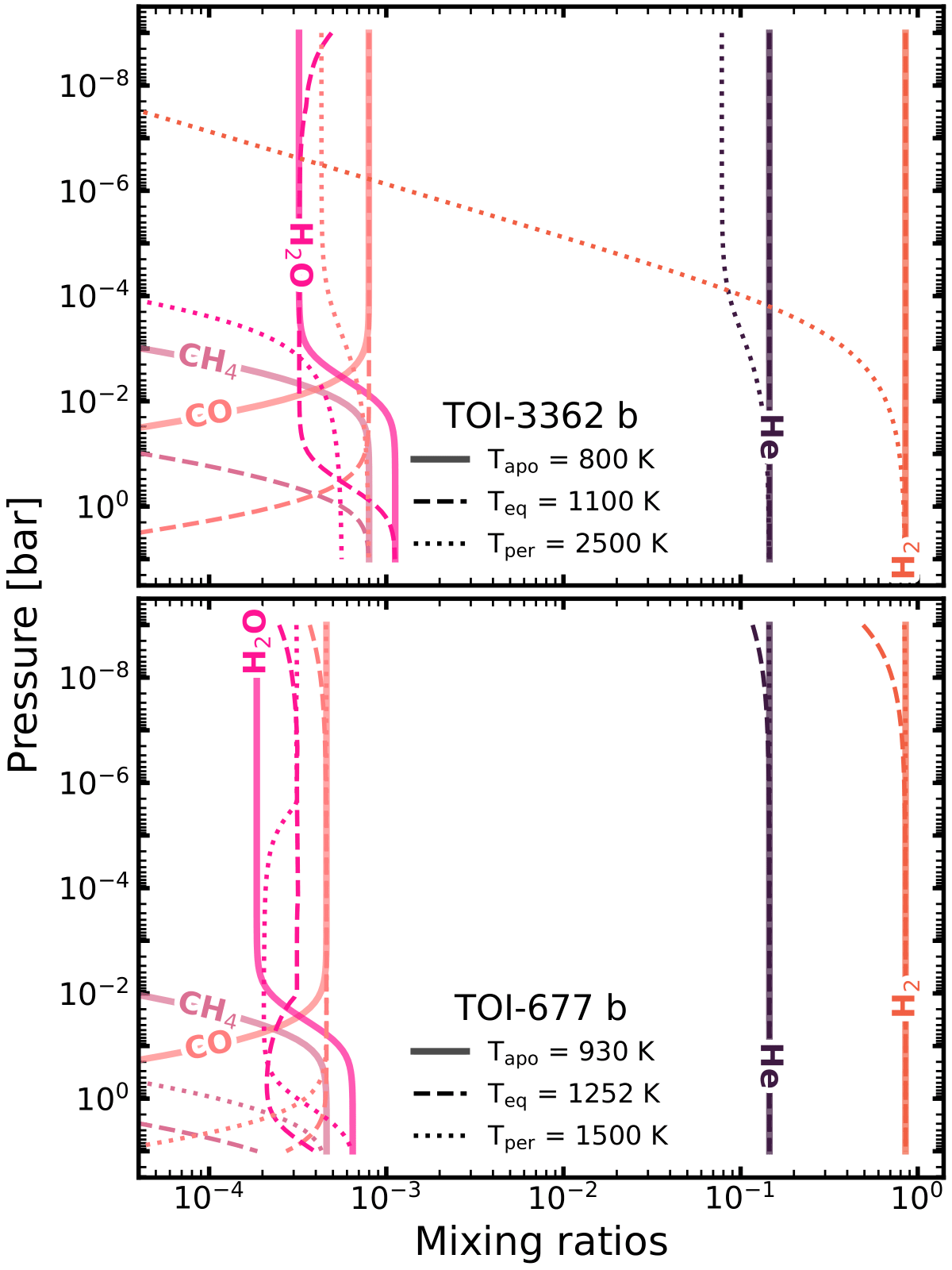}
    \caption{\textit{Top:} Predicted volume mixing ratios for \toi at the expected minimum temperature at apoastron $T_{\rm apo} = \SI{800}{\kelvin}$ (solid), the equilibrium temperature calculated assuming an albedo of 0 $T_{\rm eq} = \SI{1100}{\kelvin}$ (dotted), and the expected maximum temperature at periastron $T_{\rm per} = \SI{2500}{\kelvin}$ (dashed) \citep[see][for temperature range]{espinoza-retamal_aligned_2023}.  \textit{Bottom:} Predicted volume mixing ratios for \toie at the calculated minimum temperature at apoastron $T_{\rm apo} = \SI{930}{\kelvin}$ (solid), the equilibrium temperature $T_{\rm eq} = \SI{1252}{\kelvin}$  \citep[dotted,][]{jordan_toi_677_2020}, and the calculated maximum temperature at periastron $T_{\rm per} = \SI{2500}{\kelvin}$ (dashed). Note that at several locations the solid lines overlap with the dashed lines.} 
    \label{fig:fc_rainout}
\end{figure}

\subsection{Cross-correlation analysis}

We used the cross-correlation technique \citep{snellen_orbital_2010} to search for \water absorption in the atmospheres of these planets, following the methodology of \citet{hoeijmakers_high-resolution_2020}, for example. We corrected the reduced spectra for telluric contamination using \texttt{molecfit} \citep[v4.3.1][]{smette_molecfit_2015,kausch_molecfit_2015} by selecting regions with strong absorption of \water and \ch{O2} originating from Earth's atmosphere, while excluding any stellar contribution. For each exposure, the telluric model was computed separately to account for changing observational conditions. The models were then interpolated onto the same wavelength grid as the reduced data and divided out. We then corrected the spectra for two velocity components; namely the systemic velocity, as well as the stellar reflex motion due to the orbiting planet\footnote{We note that the spectra provided by the ESPRESSO DRS (v3.0.0) are already corrected for the velocity of Earth around the Solar System barycentre.}. 

To determine the latter, we made use of \texttt{PyAstronomy}'s \texttt{KeplerRVmodel} \citep{pya} by computing the radial velocity of the star depending on the orbital period $P$ of the planet, the radial velocity semi-amplitude $K$, the orbital eccentricity $\epsilon$, the time and argument of periastron $T_{\rm per}$ and $\omega$, the mass of the star $M_\ast$, the systemic velocity $v_{\rm sys}$ (which we set equal to 0 to only compute the reflex motion of the planet), the semi-major axis $a$ and the minimum mass of the planet $M_{\rm p} \sin{i}$, where $i$ is the orbital inclination. The time of periastron $T_{\rm per}$ was determined using \texttt{radvel}'s \texttt{timetrans\_to\_timeperi} \citep{fulton_radvel_2018} based on the transit centre time $T_0$, the orbital period $P$, the orbital eccentricity $\epsilon$ and the argument of periastron $\omega$. 

Correcting for these two velocity components, the systemic velocity and the reflex motion of the star due to the planet, effectively moves the spectra to the stellar rest frame accounting for the eccentric orbit of the planet. Unlike in previous work \citep[see e.g.][]{prinoth_titanium_2022,prinoth_time-resolved_2023}, we perform the velocity corrections outside of the cross-correlation analysis cascade, because the current implementation of \texttt{tayph} \citep{bibiana_prinoth_2024_11506199} only accounts for circular orbits. Once in the rest frame of the star, we initiated the cross-correlation cascade with \texttt{tayph} opting for outlier rejection, colour-correction, and manual masking of telluric residuals of deep lines (50\% and deeper), in particular in the region of strong \ch{O2} absorption bands, and any visible residuals in the time-average due to imperfect correction. For the outlier rejection, we applied an order-by-order sigma clipping algorithm that computes a running median absolute deviation over sub-bands of the time series. In each spectral order, the median absolute deviation was calculated spanning 40 pixels in width, and running over the entire order. Any pixel with a deviation larger than $5\sigma$ was rejected and interpolated. We normalised every order to a common flux level, by colour-correcting the order using a polynomial of order 3 for each exposure, accounting for time-dependent flux variations in the broad-band continuum.

For each planet, we generated individual cross-correlation templates for \water \citep[HITEMP,][]{rothman_hitran2012_2013}, as well as for \ch{CO} \citep[HITEMP,][]{rothman_hitran2012_2013} and \methane \citep[ExoMol,][]{yurchenko_exomol_2014} for \toi and \toie using \texttt{petitRADTRANS} \citep{molliere_petitradtrans_2019}. These templates assumed abundance profiles computed with \texttt{FastChem Cond} \citep{stock_fastchem_2018, stock_fastchem_2022, kitzmann_fastchem_2023}, accounting for condensation via rainout, an approximation commonly used for brown dwarf and exoplanet atmospheres \citep{marley_clouds_2013}. We adopted an isothermal temperature profile and the metallicity of the host star. Additionally, each template included continuum absorption by \ch{H2}-He and \ch{H2}-\ch{H2}. The reference pressure at the bottom of the atmosphere was chosen to be 1\,bar. Although clouds and hazes may mute some atmospheric features at these temperatures, we have opted to neglect clouds for this study, as the proper treatment of such complex effects is out-of-scope for this work, leaving their exploration to future studies. However, it must be noted that this choice comes with the caveat that the true planetary absorption lines are perhaps more muted than what is modelled here, and consequently the prediction of detection could be marginally an overestimation. The cross-correlation templates for \water are depicted in Figure \ref{fig:cc}, while for \toi and \toie, templates for \methane and \ch{CO} are illustrated in Figure \ref{fig:models}. 

Subsequently, we broadened these templates to approximately match the line-spread function of the ESPRESSO spectrograph, with a full-width at half-maximum of \SI{2.14}{\km\per\second} and utilised them to perform the cross-correlation analysis for each planet, by computing cross-correlation coefficients as in \citet{prinoth_titanium_2022} over a velocity range from -\SI{1000}{\km\per\second} to \SI{1000}{\km\per\second} in steps of \SI{1}{\km\per\second}. At the end of the cross-correlation cascade, we divided out the mean out-of-transit cross-correlation function to remove the stellar component, and applied a Gaussian high-pass filter with a width of \SI{50}{\km\per\second} to remove any residual broadband structure in the spectral direction. To compute the significance of the detections, we moved the two-dimensional cross-correlation maps into the rest frame of the planet and averaged the in-transit exposures. The significance is then computed by fitting a Gaussian to the signal at the expected location and dividing the amplitude by the standard deviation of the data away from any stellar or planetary signal or telluric residuals.

To determine whether the planetary absorption features are isolated from the stellar lines, we estimated the expected radial velocity extent of the RM effect by projecting the planet's position onto the stellar disk, similar to \citet{prinoth_atlas_2024}, but for eccentric orbits. 
Following \citet{sedaghati_orbital_2023}, we determined the orbital parameters through the modelling of the RM effect for all systems, the results for which are presented in Table\,\ref{tab:planets}. The planet's position in the orbital plane is given as follows:

\begin{equation}
    \mathbf{r}_{\rm op} = 
    \begin{bmatrix}
    x_{\rm op} \\
    y_{\rm op} \\
    z_{\rm op} 
    \end{bmatrix} = \begin{bmatrix}
    \frac{a}{R_\ast} \left( \cos{E} - \epsilon  \right) \\
    \frac{a}{R_\ast} \sqrt{1 - \epsilon^2} \sin{E} \\
    0 
    \end{bmatrix},
\end{equation}


\noindent where $a/R_\ast$ is the scaled semi-major axis, $E$ is the eccentric anomaly derived via Kepler's equation, and $\epsilon$ is the eccentricity of the orbit. The planetary orbit is then rotated towards the observer using the argument of periastron $\omega$ as follows:

\begin{align}
    \mathbf{r}_{\rm tp} = 
    \begin{bmatrix}
    x_{\rm tp} \\
    y_{\rm tp} \\
    z_{\rm tp} 
    \end{bmatrix} &= \begin{bmatrix}
    \cos\left(\omega - \frac{\pi}{2}\right) &  -\sin\left(\omega - \frac{\pi}{2}\right) & 0 \\
    \sin\left(\omega - \frac{\pi}{2}\right) &  \cos\left(\omega - \frac{\pi}{2}\right) & 0 \\
    0 & 0 & 1
    \end{bmatrix} 
     \mathbf{r}_{\rm op} \\ &=
     \begin{bmatrix}
    \sin\left(\omega \right) &  \cos\left(\omega \right) & 0 \\
    -\cos\left(\omega\right) &  \sin\left(\omega \right) & 0 \\
    0 & 0 & 1
    \end{bmatrix} 
     \mathbf{r}_{\rm op} 
\end{align}


\noindent The additional angle of $\pi/2$ accounts for the definition of the argument of periastron for circular orbits, $\omega_{\rm circ} \coloneqq 90 \deg $. We account for the orbital inclination $i$ relative to the observer by projecting the coordinates into the plane of the sky by:

\begin{equation}
    \mathbf{r}_{\rm sky} = 
    \begin{bmatrix}
    x_{\rm sky} \\
    y_{\rm sky} \\
    z_{\rm sky} 
    \end{bmatrix} = \begin{bmatrix}
    0 & 1 & 0 \\
    -\cos\left(i\right) & 0 & 0 \\
    \sin\left(i\right) & 0 & 0
    \end{bmatrix} 
    \mathbf{r}_{\rm tp}
\end{equation}

\noindent and account for the projected alignment of the planet's orbital plane relative to the stellar rotation through the spin-orbit alignment angle $\lambda$ as follows:

\begin{align}
    \mathbf{r}_{\rm star} = 
    \begin{bmatrix}
    x_{\rm star} \\
    y_{\rm star} \\
    z_{\rm star} 
    \end{bmatrix} = \begin{bmatrix}
    \cos(\lambda) & -\sin(\lambda) & 0 \\
    \sin(\lambda) & \cos(\lambda) & 0 \\
    0 & 0 & 1
    \end{bmatrix} 
    \mathbf{r}_{\rm sky},
\end{align}


\noindent Assuming no differential rotation, the stellar radial velocity extent of the portion behind the planet is then given by:
\begin{align}
    v_{\rm RM, RV} = x_{\rm star} v \sin{I_\ast},
    \label{eq:RV_RM}
\end{align}
where $v \sin{I_\ast}$ is the projected rotational velocity of the host star (see \citet{cegla_rossiter-mclaughlin_2016} for the calculation of the circular case). This calculation enables the determination of the radial velocity extent of the RM effect, facilitating the estimation of whether the planetary signature may be contaminated by its overlap. This assessment is particularly valuable for identifying potential contamination of the planetary signal, even if the RM effect is not directly observable in the cross-correlation map. We provide our code for the calculation of the expected radial velocity extents for the RM effect, as well as the stellar and planetary velocities, and residual telluric contamination in \citet{bibiana_prinoth_RV} and discuss its functionalities in Appendix\,\ref{app:traces}. This resource enables more careful planning of observations, particularly regarding telluric contamination.

\subsection{Model injection}
\label{sec:model_injection}

In addition, we also modelled the atmospheric spectra of \toi and \toie under the same assumptions as the cross-correlation templates including the absorption of \water, \methane and \ch{CO}, as well as molecular hydrogen (\ch{H2}) and helium {\ch{He}}, as these species are predicted to be the dominant absorbers, see Fig. \ref{fig:fc_rainout}. At lower temperatures, the atmospheres are dominated by \ch{H2} and He, with \water and CO being the next most abundant species at higher altitudes. Below approximately 0.01 bar, CO becomes less abundant and \methane becomes the dominant species instead. At \toi's periastron, where the temperature reaches ultra-hot Jupiter regimes ($>$ \SI{2000}{\kelvin}), \ch{H2}, \methane and CO start to dissociate into their atomic components at all altitudes, leading to mixing ratios below the considered range. While \water also starts to dissociate, it is still expected to be present at lower altitudes. We created four models for each planet by multiplying the mass fraction of \water by factors of 1 (nominal), 10, 100 and 1000, as illustrated in Fig. \ref{fig:models}. The models were broadened to match the resolution of the spectrograph, similar to the cross-correlation templates, but also with respect to the expected planetary rotation due to tidal locking.

We focused on these two planets due to their favourable orbital configurations, particularly the argument of periastron ($\omega$), which along with the eccentricity, ensures that the planet has a significant radial velocity, in the stellar rest frame, during transit. Additionally, at temperatures below $\sim$\SI{1000}{\kelvin}, chemical reaction timescales begin to increase exponentially. This implies that for planets with longer orbital periods and consequently cooler equilibrium temperatures, the equilibrium chemistry modelled by \texttt{FastChem Cond} may never be reached, as the chemical timescale exceeds the orbital period and vertical mixing timescales, and quenching becomes important \citep[e.g.][]{visscher_quenching_2011,visscher_chemical_2012}.

We injected the models into the raw data of \toi and \toie at the expected velocities of the planets by normalising the models to 1 and multiplying the transmission spectra \citep[see][]{prinoth_titanium_2022}. The planetary reflex motion was calculated using Kepler's third law of planetary motion based on the reflex motion of the star ($v_\ast$). The planet's radial velocity is then estimated through the conservation of angular momentum:
\begin{align}
    v_{\rm p} = - v_\ast \frac{M_\ast}{M_{\rm p}},
    \label{eq:planet_RFM}
\end{align}
where $M_\ast$ and $M_{\rm p}$ are the masses of the star and the planet, respectively. After injecting the model, we performed the same cross-correlation analysis as detailed above, searching for the injected signal of the atmosphere.

\begin{figure*}[ht!]
    \centering
    \includegraphics[width=\linewidth]{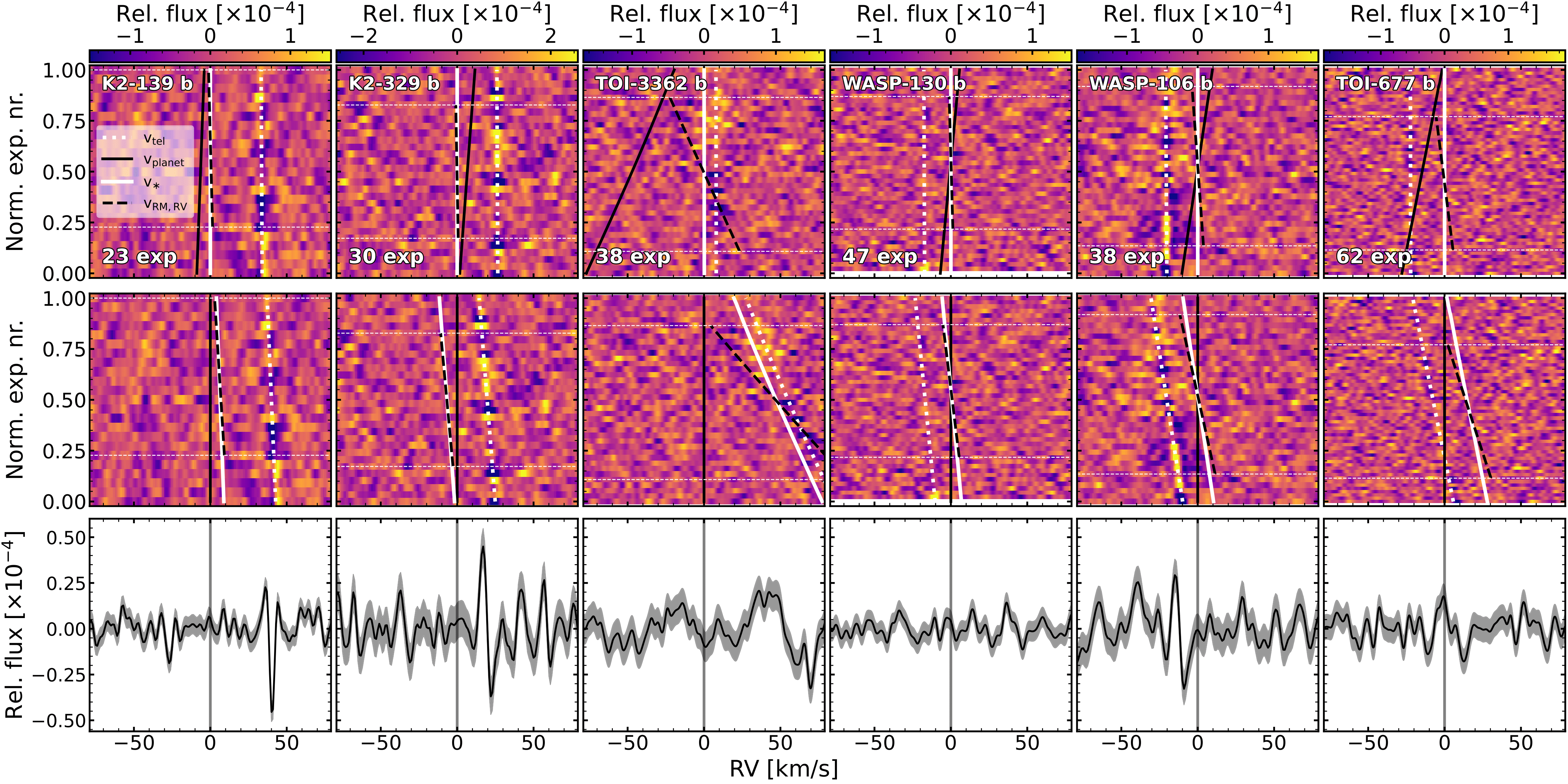}
    \caption{Water cross-correlation results for \allplanets (from left to right). \textit{Top panels:} Two-dimensional cross-correlation function in the stellar rest frame as a function of radial velocity versus normalised exposure number. We normalised the exposure number for plotting purposes, but indicated the number of exposures in the panel. The radial velocities of the telluric residuals $v_{\rm tel}$ (dotted, white), the planet $v_{\rm planet}$ (solid, black), the star $v_{\rm star}$ (solid, white) and the RM effect $v_{\rm RM, RV}$ (dashed, black) are shown. The contact times of the transit, ingress and egress are indicated with white horizontal dashed lines. Note that we masked out the first exposure of the transit of WASP-130\,b due to telluric contamination. \textit{Middle panels:} Same as top panels but in the rest frame of the planet. \textit{Bottom panels:} One-dimensional cross-correlation function, given as the time-average (vertical averaging) of the middle panels.}
    \label{fig:results}
\end{figure*}

\subsection{Simulated observations with ANDES}

To investigate the prospects for using ANDES to observe atmospheres of warm Jupiters, we simulated observations for \toi and \toie, covering its goal wavelength range up to \SI{2400}{\nm} using version 1.1 of the Exposure Time Calculator (ETC) \footnote{\url{http://tirgo.arcetri.inaf.it/nicoletta/etc_andes_sn_com.html}}. The ETC calculates S/N at a single wavelength, so we interpolated between the centres of the B, V, J, H, and K bands, where the magnitudes of the host star are known. The expected resolving power of ANDES is $\mathcal{R} \sim 100,000$, comparable to that of CRIRES+. To estimate the pixel sampling, which also includes the oversampling of the resolution element, we calculated the average spacing between neighbouring wavelengths for the Y, J, H, and K bands of CRIRES+ with 2048 pixels per order, as provided in the User Manual\footnote{\url{https://www.eso.org/sci/facilities/paranal/instruments/crires/doc/CRIRES_User_Manual_P114.1.png}, page 87, section 7.2}. This yielded an average spacing of 0.01\,nm between two neighbouring wavelength points.

We also assumed exposure times of 310 seconds for \toi and 180 seconds for \toie, consistent with the observations in this study (see Table,\ref{tab:observation_log}), resulting in average S/N at the band centres of 490 and 640 per exposure, respectively (compared to 39 and 50 with ESPRESSO). Following the approach of \citet{lee_mantis_2022}, we simulated observations with a one-hour baseline, consisting of half an hour before and after transit. This totalled 4 hours and 3.5 hours of observations, respectively, factoring in overheads due to readout, set to \SI{70}{\second}, akin to ESPRESSO's 2x1 binning mode. Although we have chosen exposure times as were chosen for the ESPRESSO observations, in reality with ANDES one would choose much shorter values in order to more finely sample the transit and thereby avoid smearing of the atmospheric signal, while maintaining a high enough S/N \citep{boldt-christmas_optimising_2023}.

To generate the synthetic observations, we followed the procedure outlined in \citet{lee_mantis_2022}, Section 7.2.2. The atmospheric spectra of the planets were assumed to be in the nominal case (1x water fractions), and the star was modelled using a PHOENIX spectrum, adopting the stellar parameters listed in Table\,\ref{tab:planets}. After correcting for the Keplerian velocities, the combined observed spectra were created by multiplying the stellar spectra with the planetary transmission spectra and then decomposed into a set of echelle orders to imitate the true use-case and Gaussian noise was added based on the calculated S/N. We then conducted the cross-correlation analysis as outlined above to predict the expected detectability of \water, \ch{CO} and \methane absorption with ANDES for \toi and \toie, assuming perfect correction for tellurics and no residual stellar contamination. This simulator that links \texttt{tayph}, \texttt{FastChem Cond}, and \texttt{petitRADTRANS} together is published in \citet{bibiana_prinoth_ExoSim}.

\section{Results and discussion}
\label{sec:results}

\begin{figure*}
    \centering
    \includegraphics[width=0.8\linewidth]{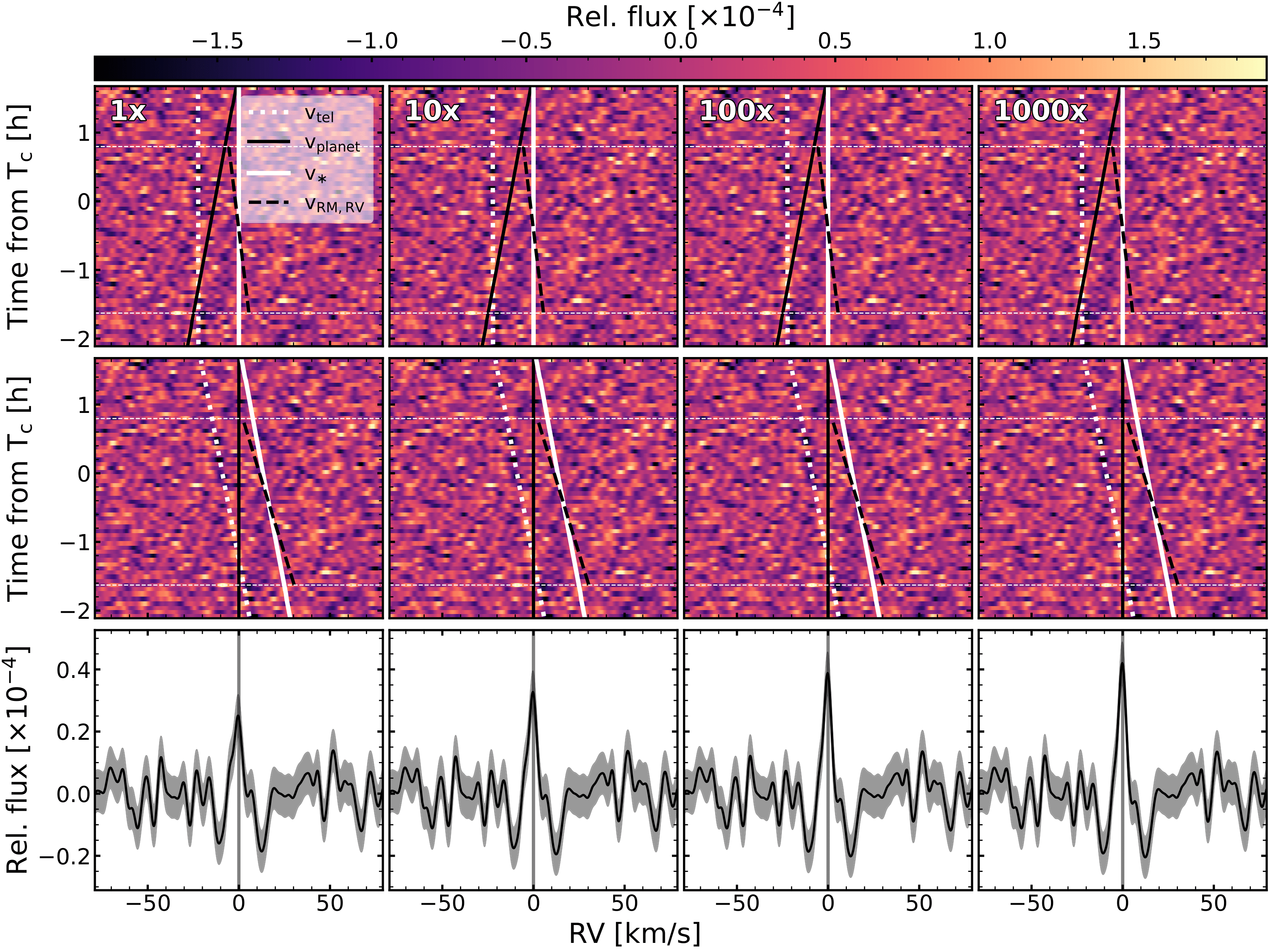}
    \caption{Cross-correlation results for model injection in the wavelength range of ESPRESSO of \toie using 1x (nominal), 10x, 100x and 1000x the predicted \water abundance, see Fig.\,\ref{fig:models} for the model spectra. Panels as in Fig.\,\ref{fig:results}.} 
    \label{fig:inject_toie}
\end{figure*}

\begin{figure*}
    \centering
    \includegraphics[width=0.8\linewidth]{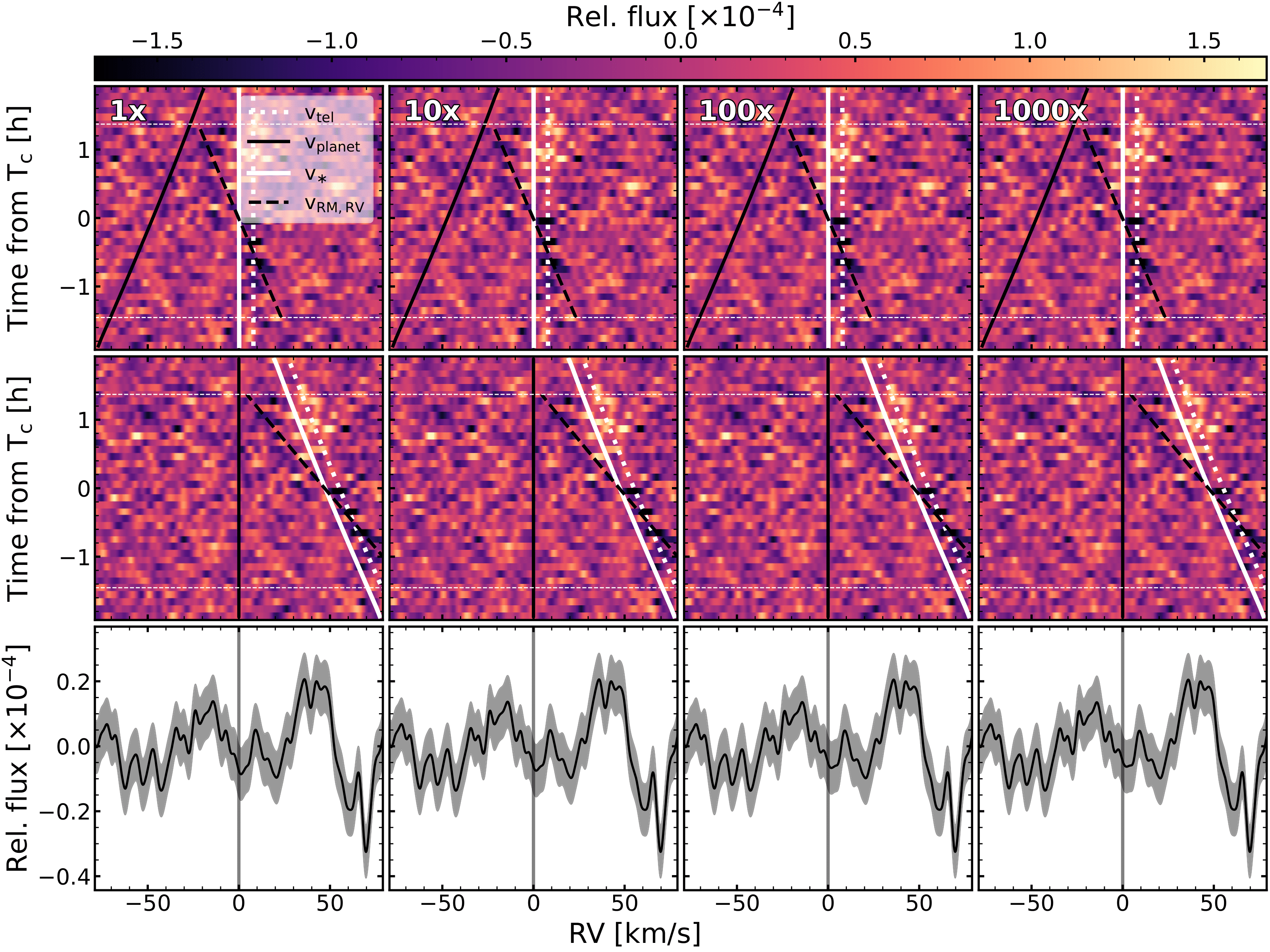}
    \caption{Same as Fig.\,\ref{fig:inject_toie} but for \toi. We note that the peak to the right, around \SI{40}{\km\per\second} is likely caused by a combination of telluric and stellar residuals. 
    }
    \label{fig:inject_toi}
\end{figure*}

\begin{figure*}
    \centering
    \includegraphics[width=0.6\textwidth]{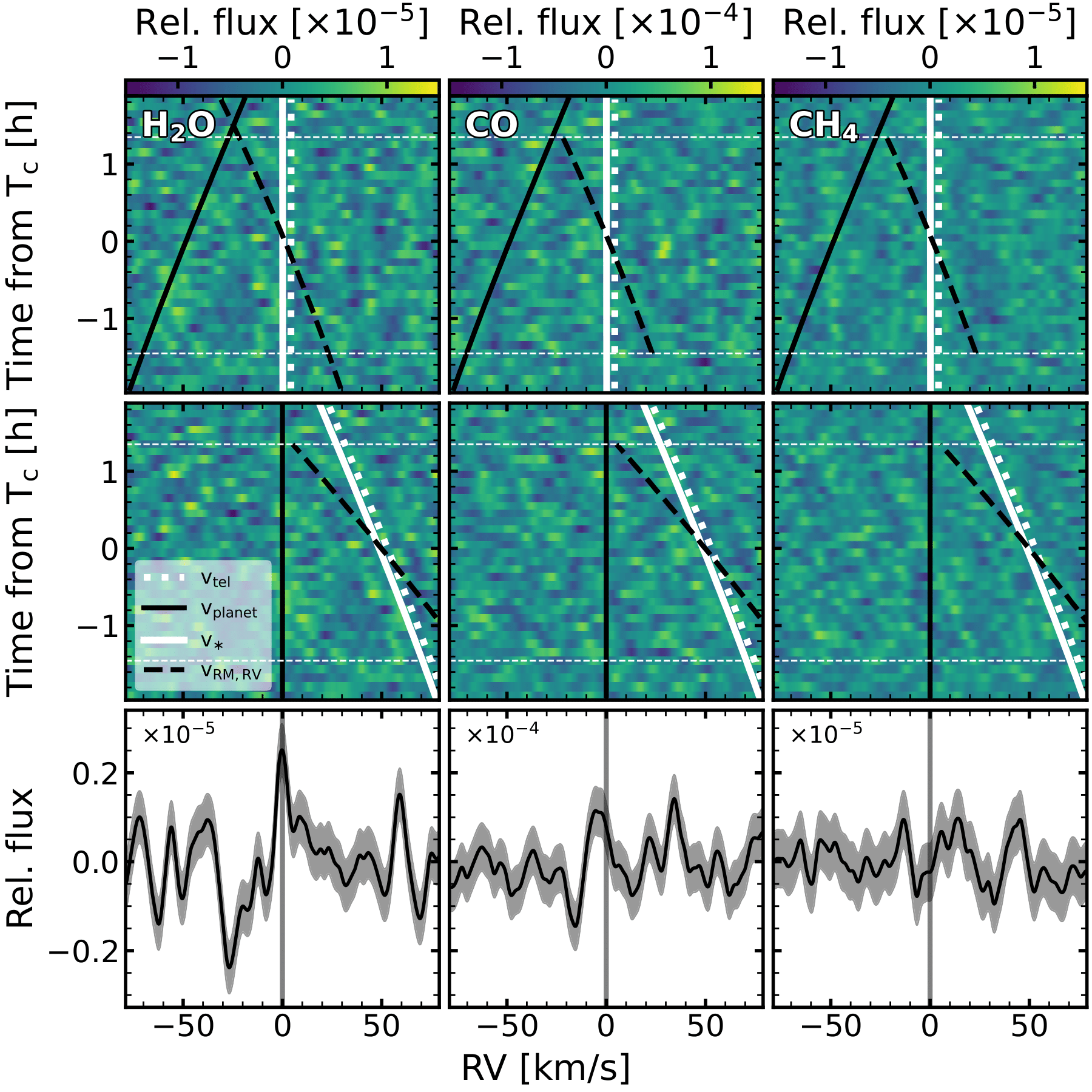}
    \caption{Same as Fig.\,\ref{fig:inject_toi} but for the simulated observations with ANDES of \toi. 
    Note the different scales for the different species.}
    \label{fig:result_ANDES_toi}
\end{figure*}

\begin{figure*}
    \centering
    \includegraphics[width=0.6\linewidth]{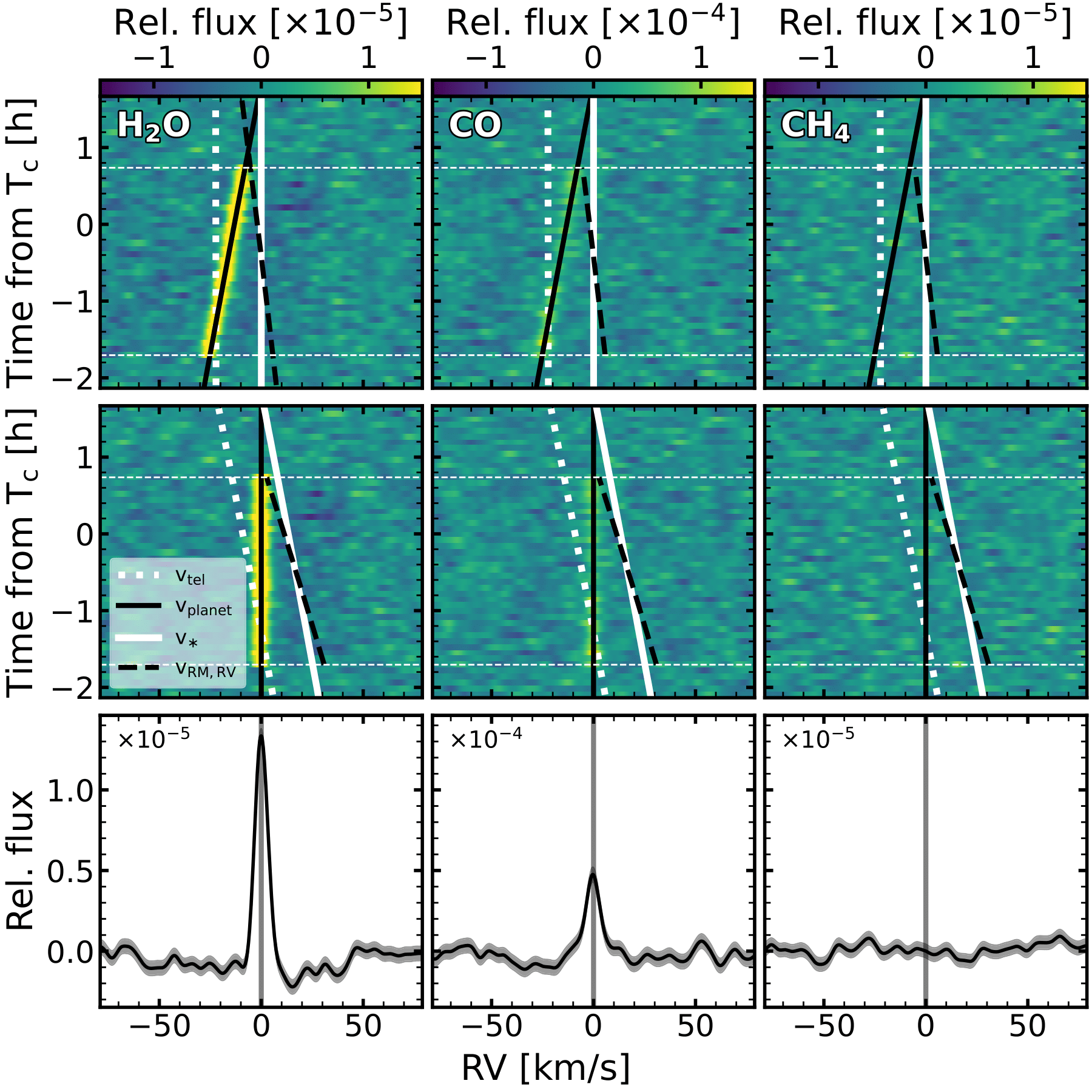}
    \caption{Same as Fig.\,\ref{fig:result_ANDES_toi} but for \toie.}
    \label{fig:result_ANDES_toie}
\end{figure*}

Fig.\,\ref{fig:results} shows the results of the cross-correlation analysis searching for \water for \allplanets. No \water absorption is detected in any of the planets examined in this study.

While slight enhancements could potentially be made to the \texttt{molecfit} corrections, the precision remains constrained by the S/N of the spectra, which tends to be relatively low as shown in Tab.\,\ref{tab:observation_log}, and the star's spectral type. Because these are typically slowly rotating late-type stars, it may be challenging to identify isolated telluric plus continuum regions. ANDES will circumvent the limitations of S/N, utilising ELT's large collecting area. Below we provide estimates for the capabilities of ANDES in detecting molecular species. A further limitation of ESPRESSO in detecting \water is its wavelength coverage, which again is remedied in ANDES with its extension into the near-infrared, possibly going as far as the K-band.


A further limitation inherent to the cross-correlation technique, is that it is restricted to planets spanning a fast radial velocity change during transit relative to the stellar rest frame, the stellar contribution through the RM effect and the expected telluric residual contamination compared to that of the planetary atmosphere. Specifically, this limitation implies that targets in this study residing on nearly circular orbits, where the radial velocities of the host star and the planet significantly overlap, are not considered optimal for studying atmospheres using this technique. 
With the known Earth barycentric velocities calculated by the ESPRESSO pipeline, the systemic velocities and the stellar and planetary reflex motions described by Eq.\,\eqref{eq:planet_RFM}, and the radial velocity extents of the RM effect described in Eq.\,\eqref{eq:RV_RM}, all velocity components can be determined and used to predict the observability of an atmosphere, provided that the orbital configurations are known with sufficient accuracy and precision. Fig.\,\ref{fig:results} illustrates the radial velocity extents corresponding to the star, the planet, the Rossiter-McLaughlin (RM) effect and the telluric residuals for \allplanets. 
The cross-correlation technique further operates under the assumption that one can effectively eliminate the stellar spectrum, assumed to be constant, through out-of-transit baseline observations. However, deviations from a perfectly constant star, such as the RM effect, introduce contamination that can affect the detection of planetary absorption features. Hence, planets on eccentric orbits provide an opportunity to isolate absorption features, provided their eccentricity and argument of periastron place them within a favourable radial velocity regime, as depicted in Fig.\,\ref{fig:graphics}. Particularly, highly eccentric planets with favourable arguments of periastron exhibit significant radial velocity changes during transit. Nonetheless, this comes with the caveat of shorter transit durations \citep[see Figs. 3.3.1 and 3.3.2 in][]{carter_estimation_2019}, thereby limiting the number of high S/N observations that can be obtained during transit.





Furthermore, observations can be strategically scheduled to ensure that telluric residuals remain distinct from the planetary velocity, as it is feasible to calculate the barycentric Earth velocity for a given time. Hence, in future studies of the atmospheres of longer-period planets, emphasis should be placed on a clear distinction of telluric residuals from the signal both by proposing astronomers and observatories/time allocation committees alike to maximise the science return. An example of such planning was highlighted by \citet{Orell-Miquel2022} in the detection of helium triplet in the atmosphere of GJ 1214\,b. Favourable configurations for observations are given for \toi and \toie, as illustrated in Fig.\,\ref{fig:results}, where the planetary radial velocity extent is notably different from that of the star, the RM effect and any potential telluric residuals (slightly less so for \toie). While we did not detect \water in the atmospheres of \toi and \toie, our model injection shown in Figs.\,\ref{fig:inject_toie} predict the detection of \water, for the 100x case (5.75$\sigma$) and tentatively (4.8$\sigma$) for the 10x case for \toie. We note that our model assumptions do not account for clouds or hazes, even though these are to be expected at these temperatures, which means that it is likely that if \water is indeed present on \toie, these features may very well be muted. 


Fig.\,\ref{fig:inject_toi} shows the results of model injection for 1x, 10x, 100x, and 1000x the nominal \water abundance in the case of \toi within the ESPRESSO wavelength range. It is evident that even with the increased \water abundance, these observations do not facilitate detection. The reason lies in the nature of the relatively shallower \water bands in the optical wavelength regime \citep[see previous water searches in e.g.][]{allart_search_2017, pelletier_where_2021,allart_wasp-127b_2020}, as well as the relatively high surface gravity resulting from the substantial planetary mass of \SI{4.0 \pm 0.4}{M_{\rm Jup}} \citep{espinoza-retamal_aligned_2023}.

Figs.\,\ref{fig:result_ANDES_toi} and \ref{fig:result_ANDES_toie} show the results of the simulated observations with ANDES using the nominal model of \toi and \toie, respectively. The simulated observations spanned the goal wavelength range up to 2400\,nm, which includes the K-band that provides access to the prominent \ch{CO} feature around 2300\,nm. Despite the high surface gravity of \toi, our simulated observations predict a tentative detection of \water at a relative absorption depth of $2.13 \pm 0.32$ ppm (4.0$\sigma$), leveraging its ability to probe deeper \water bands, although still shallower than those of other planets in the sample. On the other hand, \ch{CO} detection is not predicted.

In contrast, our simulated observations for \toie predict robust detections for both \water and \ch{CO} with absorption depths of $13.8 \pm 0.4$ ppm (31.0$\sigma$) and $44.7 \pm 3.1$ ppm (10.4$\sigma$), respectively, as illustrated in Fig.,\ref{fig:result_ANDES_toie}. Although the orbital configuration appears less optimal than that of \toi at first glance, it is worth noting that future observations could optimise the velocity extent of the telluric contamination by planning observation windows accordingly. \methane detection is not predicted as it remains confined to the lower layers of the atmospheres for both models, see Fig.\,\ref{fig:fc_rainout}, where transmission spectroscopy lacks sensitivity.

The \water bands covered by ESPRESSO are relatively shallow ($\sim$0.5 ppm transit depth), probing lower regions of the atmosphere where transmission spectroscopy is less sensitive. In turn, towards redder wavelengths, i.e. the wavelengths where ANDES will be, and CRIRES+ already is, probing, the \water bands are deeper, providing access to higher altitudes of the atmosphere. For targets with favourable configurations and planetary parameters, current and upcoming infrared instrumentation are expected to be the preferred choice for observing warm Jupiters on eccentric orbits to investigate the absorption of \water.

The results of these simulations showcase one of the primary scientific goals of the ANDES instrument at the ELT in detecting exoplanetary atmospheres, where it will open up the parameter space of detection, significantly pushing the boundaries of which kind of planets are accessible for atmospheric studies. 



\section{Conclusions}
\label{sec:Conclusions}

In this study, we investigate the capabilities and limitations of current and upcoming instrumentation for detecting atmospheric features of exoplanets with orbital periods exceeding 10 days, using the cross-correlation technique. Our analysis aims to detect the absorption of \water in the transmission spectra of \allplanets using ESPRESSO observations, and investigate predictions for the ELT instrument ANDES for \toi and \toie due to their favourable orbital configurations.

While our cross-correlation analysis using ESPRESSO data does not detect the presence of \water absorption in any of the observed planets, primarily due to insufficient radial velocity change and the extent of the planets relative to the star, our findings underscore the challenges of disentangling signals in systems with circular orbits. For planets like K2-139\,b, K2-329\,b, WASP-106\,b and WASP-130\,b on (nearly) circular orbits, overlapping radial velocities between the star and the planet, together with small radial velocity changes due to the large orbital distance, complicate the detection of atmospheric features, resulting in contamination by stellar residuals, the RM effect or telluric residuals.

Despite these challenges, our model injections for planets on eccentric orbits, \toi and \toie, suggest the potential to detect \water in \toie. Conversely, due to the large surface gravity of \toi, no detection is predicted within the wavelength range covered by ESPRESSO. However, it is important to note that atmospheric features may be attenuated by the presence of clouds and hazes in the atmospheres of such planets.

As part of our study, we present a simulation tool tailored for upcoming ANDES observations, allowing us to assess the detectability of atmospheric features. Using this tool, we simulated observations of the two planets with favourable orbital configurations, \toi and \toie. Our simulations yield promising results, predicting significant detections of \water in the atmospheres of both planets, along with the detection of \ch{CO} for \toie, if ANDES indeed covers the K-band. These findings provide valuable insights into the capabilities of ANDES for cross-correlation studies of exoplanetary atmospheres with orbits beyond 10 days and highlight the importance of prioritising planets with favourable orbital configurations for future observational campaigns.

In the coming years, careful selection and characterisation of warm Jupiter-like planets with favourable orbital configurations will be crucial in preparation for ANDES. By doing so, we can maximise the chances of detecting atmospheric signatures using the cross-correlation technique for colder gas giants, advancing our understanding of more diverse exoplanetary atmospheres.

\section*{Acknowledgements}
This work is based on observations collected at the European Southern Observatory under ESO programmes 109.238M, 108.22C0, and 110.23Y8. This research has made use of the services of the ESO Science Archive Facility. The authors thank the ESPRESSO team for building and maintaining the instrument. This research has made use of the NASA Exoplanet Archive, which is operated by the California Institute of Technology, under contract with the National Aeronautics and Space Administration under the Exoplanet Exploration Program. The authors thank Daniel Kitzmann for helping us understand \texttt{FastChem Cond}. This study makes use of \texttt{astropy} \citep{astropy:2013,astropy:2018,astropy:2022} and {\tt label-lines} \citep{cadiou_matplotlib_2022}.
B.P.\ acknowledges financial support from The Fund of the Walter Gyllenberg Foundation. B.T.\ acknowledges the financial support from the Wenner-Gren Foundation (WGF2022-0041). A.J.\ and R.B.\ acknowledge support from ANID -- Millennium  Science Initiative -- ICN12\_009.  R.B.\ acknowledges support from FONDECYT Project 1241963. A.J.\ acknowledges support from FONDECYT project 1210718. We also would like to thank the anonymous referee for their comments and suggestions that helped improve the quality of the manuscript.


\begin{thebibliography}{}
\expandafter\ifx\csname natexlab\endcsname\relax\def\natexlab#1{#1}\fi
\providecommand{\url}[1]{\href{#1}{#1}}
\providecommand{\dodoi}[1]{doi:~\href{http://doi.org/#1}{\nolinkurl{#1}}}
\providecommand{\doeprint}[1]{\href{http://ascl.net/#1}{\nolinkurl{http://ascl.net/#1}}}
\providecommand{\doarXiv}[1]{\href{https://arxiv.org/abs/#1}{\nolinkurl{https://arxiv.org/abs/#1}}}

\bibitem[{Ahrer {et~al.}(2022)Ahrer, Stevenson, Mansfield, Moran, Brande, Morello, Murray, Nikolov, de~la Roche, Schlawin, Wheatley, Zieba, Batalha, Damiano, Goyal, Lendl, Lothringer, Mukherjee, Ohno, Batalha, Battley, Bean, Beatty, Benneke, Berta-Thompson, Carter, Cubillos, Daylan, Espinoza, Gao, Gibson, Gill, Harrington, Hu, Kreidberg, Lewis, Line, López-Morales, Parmentier, Powell, Sing, Tsai, Wakeford, Welbanks, Alam, Alderson, Allen, Anderson, Barstow, Bayliss, Bell, Blecic, Bryant, Burleigh, Carone, Casewell, Changeat, Chubb, Crossfield, Crouzet, Decin, Désert, Feinstein, Flagg, Fortney, Gizis, Heng, Iro, Kempton, Kendrew, Kirk, Knutson, Komacek, Lagage, Leconte, Lustig-Yaeger, MacDonald, Mancini, May, Mayne, Miguel, Mikal-Evans, Molaverdikhani, Palle, Piaulet, Rackham, Redfield, Rogers, Roy, Rustamkulov, Shkolnik, Sotzen, Taylor, Tremblin, Tucker, Turner, de~Val-Borro, Venot, \& Zhang}]{ahrer_early_2022}
Ahrer, E.-M., Stevenson, K.~B., Mansfield, M., {et~al.} 2022, Early {Release} {Science} of the exoplanet {WASP}-39b with {JWST} {NIRCam},  arXiv, \dodoi{10.48550/arXiv.2211.10489}

\bibitem[{Alderson {et~al.}(2022)Alderson, Wakeford, Alam, Batalha, Lothringer, Redai, Barat, Brande, Damiano, Daylan, Espinoza, Flagg, Goyal, Grant, Hu, Inglis, Lee, Mikal-Evans, Ramos-Rosado, Roy, Wallack, Batalha, Bean, Benneke, Berta-Thompson, Carter, Changeat, Colón, Crossfield, Désert, Foreman-Mackey, Gibson, Kreidberg, Line, López-Morales, Molaverdikhani, Moran, Morello, Moses, Mukherjee, Schlawin, Sing, Stevenson, Taylor, Aggarwal, Ahrer, Allen, Barstow, Bell, Blecic, Casewell, Chubb, Crouzet, Cubillos, Decin, Feinstein, Fortney, Harrington, Heng, Iro, Kempton, Kirk, Knutson, Krick, Leconte, Lendl, MacDonald, Mancini, Mansfield, May, Mayne, Miguel, Nikolov, Ohno, Palle, Parmentier, de~la Roche, Piaulet, Powell, Rackham, Redfield, Rogers, Rustamkulov, Tan, Tremblin, Tsai, Turner, de~Val-Borro, Venot, Welbanks, Wheatley, \& Zhang}]{alderson_early_2022}
Alderson, L., Wakeford, H.~R., Alam, M.~K., {et~al.} 2022, Early {Release} {Science} of the {Exoplanet} {WASP}-39b with {JWST} {NIRSpec} {G395H},  arXiv, \dodoi{10.48550/arXiv.2211.10488}

\bibitem[{Allart {et~al.}(2017)Allart, Lovis, Pino, Wyttenbach, Ehrenreich, \& Pepe}]{allart_search_2017}
Allart, R., Lovis, C., Pino, L., {et~al.} 2017, Astronomy and Astrophysics, 606, A144, \dodoi{10.1051/0004-6361/201730814}

\bibitem[{Allart {et~al.}(2020)Allart, Pino, Lovis, Sousa, Casasayas-Barris, Zapatero~Osorio, Cretignier, Palle, Pepe, Cristiani, Rebolo, Santos, Borsa, Bourrier, Demangeon, Ehrenreich, Lavie, Lendl, Lillo-Box, Micela, Oshagh, Sozzetti, Tabernero, Adibekyan, Allende~Prieto, Alibert, Amate, Benz, Bouchy, Cabral, Dekker, D'Odorico, Di~Marcantonio, Dumusque, Figueira, Genova~Santos, González~Hernández, Lo~Curto, Manescau, Martins, Mégevand, Mehner, Molaro, Nunes, Poretti, Riva, Suárez~Mascareño, Udry, \& Zerbi}]{allart_wasp-127b_2020}
Allart, R., Pino, L., Lovis, C., {et~al.} 2020, A\&A, 644, A155, \dodoi{10.1051/0004-6361/202039234}

\bibitem[{{Astropy Collaboration} {et~al.}(2013){Astropy Collaboration}, {Robitaille}, {Tollerud}, {Greenfield}, {Droettboom}, {Bray}, {Aldcroft}, {Davis}, {Ginsburg}, {Price-Whelan}, {Kerzendorf}, {Conley}, {Crighton}, {Barbary}, {Muna}, {Ferguson}, {Grollier}, {Parikh}, {Nair}, {Unther}, {Deil}, {Woillez}, {Conseil}, {Kramer}, {Turner}, {Singer}, {Fox}, {Weaver}, {Zabalza}, {Edwards}, {Azalee Bostroem}, {Burke}, {Casey}, {Crawford}, {Dencheva}, {Ely}, {Jenness}, {Labrie}, {Lim}, {Pierfederici}, {Pontzen}, {Ptak}, {Refsdal}, {Servillat}, \& {Streicher}}]{astropy:2013}
{Astropy Collaboration}, {Robitaille}, T.~P., {Tollerud}, E.~J., {et~al.} 2013, \aap, 558, A33, \dodoi{10.1051/0004-6361/201322068}

\bibitem[{{Astropy Collaboration} {et~al.}(2018){Astropy Collaboration}, {Price-Whelan}, {Sip{\H{o}}cz}, {G{\"u}nther}, {Lim}, {Crawford}, {Conseil}, {Shupe}, {Craig}, {Dencheva}, {Ginsburg}, {Vand erPlas}, {Bradley}, {P{\'e}rez-Su{\'a}rez}, {de Val-Borro}, {Aldcroft}, {Cruz}, {Robitaille}, {Tollerud}, {Ardelean}, {Babej}, {Bach}, {Bachetti}, {Bakanov}, {Bamford}, {Barentsen}, {Barmby}, {Baumbach}, {Berry}, {Biscani}, {Boquien}, {Bostroem}, {Bouma}, {Brammer}, {Bray}, {Breytenbach}, {Buddelmeijer}, {Burke}, {Calderone}, {Cano Rodr{\'\i}guez}, {Cara}, {Cardoso}, {Cheedella}, {Copin}, {Corrales}, {Crichton}, {D'Avella}, {Deil}, {Depagne}, {Dietrich}, {Donath}, {Droettboom}, {Earl}, {Erben}, {Fabbro}, {Ferreira}, {Finethy}, {Fox}, {Garrison}, {Gibbons}, {Goldstein}, {Gommers}, {Greco}, {Greenfield}, {Groener}, {Grollier}, {Hagen}, {Hirst}, {Homeier}, {Horton}, {Hosseinzadeh}, {Hu}, {Hunkeler}, {Ivezi{\'c}}, {Jain}, {Jenness}, {Kanarek}, {Kendrew}, {Kern}, {Kerzendorf}, {Khvalko}, {King}, {Kirkby}, {Kulkarni},
  {Kumar}, {Lee}, {Lenz}, {Littlefair}, {Ma}, {Macleod}, {Mastropietro}, {McCully}, {Montagnac}, {Morris}, {Mueller}, {Mumford}, {Muna}, {Murphy}, {Nelson}, {Nguyen}, {Ninan}, {N{\"o}the}, {Ogaz}, {Oh}, {Parejko}, {Parley}, {Pascual}, {Patil}, {Patil}, {Plunkett}, {Prochaska}, {Rastogi}, {Reddy Janga}, {Sabater}, {Sakurikar}, {Seifert}, {Sherbert}, {Sherwood-Taylor}, {Shih}, {Sick}, {Silbiger}, {Singanamalla}, {Singer}, {Sladen}, {Sooley}, {Sornarajah}, {Streicher}, {Teuben}, {Thomas}, {Tremblay}, {Turner}, {Terr{\'o}n}, {van Kerkwijk}, {de la Vega}, {Watkins}, {Weaver}, {Whitmore}, {Woillez}, {Zabalza}, \& {Astropy Contributors}}]{astropy:2018}
{Astropy Collaboration}, {Price-Whelan}, A.~M., {Sip{\H{o}}cz}, B.~M., {et~al.} 2018, \aj, 156, 123, \dodoi{10.3847/1538-3881/aabc4f}

\bibitem[{{Astropy Collaboration} {et~al.}(2022){Astropy Collaboration}, {Price-Whelan}, {Lim}, {Earl}, {Starkman}, {Bradley}, {Shupe}, {Patil}, {Corrales}, {Brasseur}, {N{"o}the}, {Donath}, {Tollerud}, {Morris}, {Ginsburg}, {Vaher}, {Weaver}, {Tocknell}, {Jamieson}, {van Kerkwijk}, {Robitaille}, {Merry}, {Bachetti}, {G{"u}nther}, {Aldcroft}, {Alvarado-Montes}, {Archibald}, {B{'o}di}, {Bapat}, {Barentsen}, {Baz{'a}n}, {Biswas}, {Boquien}, {Burke}, {Cara}, {Cara}, {Conroy}, {Conseil}, {Craig}, {Cross}, {Cruz}, {D'Eugenio}, {Dencheva}, {Devillepoix}, {Dietrich}, {Eigenbrot}, {Erben}, {Ferreira}, {Foreman-Mackey}, {Fox}, {Freij}, {Garg}, {Geda}, {Glattly}, {Gondhalekar}, {Gordon}, {Grant}, {Greenfield}, {Groener}, {Guest}, {Gurovich}, {Handberg}, {Hart}, {Hatfield-Dodds}, {Homeier}, {Hosseinzadeh}, {Jenness}, {Jones}, {Joseph}, {Kalmbach}, {Karamehmetoglu}, {Ka{l}uszy{'n}ski}, {Kelley}, {Kern}, {Kerzendorf}, {Koch}, {Kulumani}, {Lee}, {Ly}, {Ma}, {MacBride}, {Maljaars}, {Muna}, {Murphy}, {Norman}, {O'Steen},
  {Oman}, {Pacifici}, {Pascual}, {Pascual-Granado}, {Patil}, {Perren}, {Pickering}, {Rastogi}, {Roulston}, {Ryan}, {Rykoff}, {Sabater}, {Sakurikar}, {Salgado}, {Sanghi}, {Saunders}, {Savchenko}, {Schwardt}, {Seifert-Eckert}, {Shih}, {Jain}, {Shukla}, {Sick}, {Simpson}, {Singanamalla}, {Singer}, {Singhal}, {Sinha}, {Sip{H{o}}cz}, {Spitler}, {Stansby}, {Streicher}, {{{S}}umak}, {Swinbank}, {Taranu}, {Tewary}, {Tremblay}, {Val-Borro}, {Van Kooten}, {Vasovi{'c}}, {Verma}, {de Miranda Cardoso}, {Williams}, {Wilson}, {Winkel}, {Wood-Vasey}, {Xue}, {Yoachim}, {Zhang}, {Zonca}, \& {Astropy Project Contributors}}]{astropy:2022}
{Astropy Collaboration}, {Price-Whelan}, A.~M., {Lim}, P.~L., {et~al.} 2022, ApJ, 935, 167, \dodoi{10.3847/1538-4357/ac7c74}

\bibitem[{Barragán {et~al.}(2018)Barragán, Gandolfi, Smith, Deeg, Fridlund, Persson, Donati, Endl, Csizmadia, Grziwa, Nespral, Hatzes, Cochran, Fossati, Brems, Cabrera, Cusano, Eigmüller, Eiroa, Erikson, Guenther, Korth, Lorenzo-Oliveira, Mancini, Pätzold, Prieto-Arranz, Rauer, Rebollido, Saario, \& Zakhozhay}]{barragan_k2-139_2018}
Barragán, O., Gandolfi, D., Smith, A. M.~S., {et~al.} 2018, Monthly Notices of the Royal Astronomical Society, 475, 1765, \dodoi{10.1093/mnras/stx3207}

\bibitem[{Birkby(2018)}]{birkby_exoplanet_2018}
Birkby, J.~L. 2018, Exoplanet {Atmospheres} at {High} {Spectral} {Resolution}, \dodoi{10.48550/arXiv.1806.04617}

\bibitem[{Bohn {et~al.}(2020)Bohn, Southworth, Ginski, Kenworthy, Maxted, \& Evans}]{bohn_multiplicity_2020}
Bohn, A.~J., Southworth, J., Ginski, C., {et~al.} 2020, Astronomy and Astrophysics, 635, A73, \dodoi{10.1051/0004-6361/201937127}

\bibitem[{Boldt-Christmas {et~al.}(2023)Boldt-Christmas, Lesjak, Wehrhahn, Piskunov, Rains, Nortmann, \& Kochukhov}]{boldt-christmas_optimising_2023}
Boldt-Christmas, L., Lesjak, F., Wehrhahn, A., {et~al.} 2023, Optimising spectroscopic observations of transiting exoplanets,  arXiv, \dodoi{10.48550/arXiv.2312.08320}

\bibitem[{Bouchy {et~al.}(2017)Bouchy, Doyon, Artigau, Melo, Hernandez, Wildi, Delfosse, Lovis, Figueira, Canto~Martins, González~Hernández, Thibault, Reshetov, Pepe, Santos, de~Medeiros, Rebolo, Abreu, Adibekyan, Bandy, Benz, Blind, Bohlender, Boisse, Bovay, Broeg, Brousseau, Cabral, Chazelas, Cloutier, Coelho, Conod, Cumming, Delabre, Genolet, Hagelberg, Jayawardhana, Käufl, Lafrenière, de~Castro~Leão, Malo, de~Medeiros~Martins, Matthews, Metchev, Oshagh, Ouellet, Parro, Rasilla~Piñeiro, Santos, Sarajlic, Segovia, Sordet, Udry, Valencia, Vallée, Venn, Wade, \& Saddlemyer}]{bouchy_near-infrared_2017}
Bouchy, F., Doyon, R., Artigau, {et~al.} 2017, The Messenger, 169, 21, \dodoi{10.18727/0722-6691/5034}

\bibitem[{{Brahm} {et~al.}(2017{\natexlab{a}}){Brahm}, {Jord{\'a}n}, {Hartman}, \& {Bakos}}]{zaspe}
{Brahm}, R., {Jord{\'a}n}, A., {Hartman}, J., \& {Bakos}, G. 2017{\natexlab{a}}, \mnras, 467, 971, \dodoi{10.1093/mnras/stx144}

\bibitem[{{Brahm} {et~al.}(2017{\natexlab{b}}){Brahm}, {Jord{\'a}n}, {Hartman}, \& {Bakos}}]{Brahm2017}
---. 2017{\natexlab{b}}, \mnras, 467, 971, \dodoi{10.1093/mnras/stx144}

\bibitem[{{Brahm} {et~al.}(2019){Brahm}, {Espinoza}, {Jord{\'a}n}, {Henning}, {Sarkis}, {Jones}, {D{\'\i}az}, {Jenkins}, {Vanzi}, {Zapata}, {Petrovich}, {Kossakowski}, {Rabus}, {Rojas}, \& {Torres}}]{brahm:2019}
{Brahm}, R., {Espinoza}, N., {Jord{\'a}n}, A., {et~al.} 2019, \aj, 158, 45, \dodoi{10.3847/1538-3881/ab279a}

\bibitem[{{Bressan} {et~al.}(2012){Bressan}, {Marigo}, {Girardi}, {Salasnich}, {Dal Cero}, {Rubele}, \& {Nanni}}]{parsec}
{Bressan}, A., {Marigo}, P., {Girardi}, L., {et~al.} 2012, \mnras, 427, 127, \dodoi{10.1111/j.1365-2966.2012.21948.x}

\bibitem[{Cadiou(2022)}]{cadiou_matplotlib_2022}
Cadiou, C. 2022, Matplotlib label lines,  Zenodo, \dodoi{10.5281/zenodo.7428071}

\bibitem[{Carter(2019)}]{carter_estimation_2019}
Carter, J.~L. 2019, Estimation of {Planetary} {Photometric} {Emissions} for {Extremely} {Close}-in {Exoplanets}, \dodoi{10.48550/arXiv.1901.01361}

\bibitem[{Cegla {et~al.}(2016)Cegla, Lovis, Bourrier, Beeck, Watson, \& Pepe}]{cegla_rossiter-mclaughlin_2016}
Cegla, H.~M., Lovis, C., Bourrier, V., {et~al.} 2016, A\&A, 588, A127, \dodoi{10.1051/0004-6361/201527794}

\bibitem[{{Czesla} {et~al.}(2019){Czesla}, {Schr{\"o}ter}, {Schneider}, {Huber}, {Pfeifer}, {Andreasen}, \& {Zechmeister}}]{pya}
{Czesla}, S., {Schr{\"o}ter}, S., {Schneider}, C.~P., {et~al.} 2019, {PyA: Python astronomy-related packages}.
\newblock \doeprint{1906.010}

\bibitem[{Dawson \& Johnson(2018)}]{dawson_origins_2018}
Dawson, R.~I., \& Johnson, J.~A. 2018, Annual Review of Astronomy and Astrophysics, 56, 175, \dodoi{10.1146/annurev-astro-081817-051853}

\bibitem[{Donati {et~al.}(2020)Donati, Kouach, Moutou, Doyon, Delfosse, Artigau, Baratchart, Lacombe, Barrick, Hébrard, Bouchy, Saddlemyer, Parès, Rabou, Micheau, Dolon, Reshetov, Challita, Carmona, Striebig, Thibault, Martioli, Cook, Fouqué, Vermeulen, Wang, Arnold, Pepe, Boisse, Figueira, Bouvier, Ray, Feugeade, Morin, Alencar, Hobson, Castilho, Udry, Santos, Hernandez, Benedict, Vallée, Gallou, Dupieux, Larrieu, Perruchot, Sottile, Moreau, Usher, Baril, Wildi, Chazelas, Malo, Bonfils, Loop, Kerley, Wevers, Dunn, Pazder, Macdonald, Dubois, Carrié, Valentin, Henault, Yan, \& Steinmetz}]{donati_spirou_2020}
Donati, J.~F., Kouach, D., Moutou, C., {et~al.} 2020, Monthly Notices of the Royal Astronomical Society, 498, 5684, \dodoi{10.1093/mnras/staa2569}

\bibitem[{Dorn {et~al.}(2023)Dorn, Bristow, Smoker, Rodler, Lavail, Accardo, Ancker, Baade, Baruffolo, Courtney-Barrer, Blanco, Brucalassi, Cumani, Follert, Haimerl, Hatzes, Haug, Heiter, Hinterschuster, Hubin, Ives, Jung, Jones, Kaeufl, Kirchbauer, Klein, Kochukhov, Korhonen, Köhler, Lizon, Moins, Molina-Conde, Marquart, Neeser, Oliva, Pallanca, Pasquini, Paufique, Piskunov, Reiners, Schneller, Schmutzer, Seemann, Slumstrup, Smette, Stegmeier, Stempels, Tordo, Valenti, Valenzuela, Vernet, Vinther, \& Wehrhahn}]{dorn_crires_2023}
Dorn, R.~J., Bristow, P., Smoker, J.~V., {et~al.} 2023, Astronomy \& Astrophysics, 671, A24, \dodoi{10.1051/0004-6361/202245217}

\bibitem[{Espinoza-Retamal {et~al.}(2023)Espinoza-Retamal, Brahm, Petrovich, Jordán, Stefánsson, Sedaghati, Hobson, Muñoz, Boyle, Leiva, \& Suc}]{espinoza-retamal_aligned_2023}
Espinoza-Retamal, J.~I., Brahm, R., Petrovich, C., {et~al.} 2023, The Astrophysical Journal Letters, 958, L20, \dodoi{10.3847/2041-8213/ad096d}

\bibitem[{Feinstein {et~al.}(2022)Feinstein, Radica, Welbanks, Murray, Ohno, Coulombe, Espinoza, Bean, Teske, Benneke, Line, Rustamkulov, Saba, Tsiaras, Barstow, Fortney, Gao, Knutson, MacDonald, Mikal-Evans, Rackham, Taylor, Parmentier, Batalha, Berta-Thompson, Carter, Changeat, Santos, Gibson, Goyal, Kreidberg, López-Morales, Lothringer, Miguel, Molaverdikhani, Moran, Morello, Mukherjee, Sing, Stevenson, Wakeford, Ahrer, Alam, Alderson, Allen, Batalha, Bell, Blecic, Brande, Caceres, Casewell, Chubb, Crossfield, Crouzet, Cubillos, Decin, Désert, Harrington, Heng, Henning, Iro, Kempton, Kendrew, Kirk, Krick, Lagage, Lendl, Mancini, Mansfield, May, Mayne, Nikolov, Palle, de~la Roche, Piaulet, Powell, Redfield, Rogers, Roman, Roy, Nixon, Schlawin, Tan, Tremblin, Turner, Venot, Waalkes, Wheatley, \& Zhang}]{feinstein_early_2022}
Feinstein, A.~D., Radica, M., Welbanks, L., {et~al.} 2022, Early {Release} {Science} of the exoplanet {WASP}-39b with {JWST} {NIRISS},  arXiv, \dodoi{10.48550/arXiv.2211.10493}

\bibitem[{Fortney {et~al.}(2021)Fortney, Dawson, \& Komacek}]{fortney_hot_2021}
Fortney, J.~J., Dawson, R.~I., \& Komacek, T.~D. 2021, Journal of Geophysical Research (Planets), 126, \dodoi{10.1029/2020JE006629}

\bibitem[{Fortney {et~al.}(2007)Fortney, Marley, \& Barnes}]{fortney_planetary_2007}
Fortney, J.~J., Marley, M.~S., \& Barnes, J.~W. 2007, The Astrophysical Journal, 659, 1661, \dodoi{10.1086/512120}

\bibitem[{Fulton {et~al.}(2018)Fulton, Petigura, Blunt, \& Sinukoff}]{fulton_radvel_2018}
Fulton, B.~J., Petigura, E.~A., Blunt, S., \& Sinukoff, E. 2018, Publications of the Astronomical Society of the Pacific, 130, 044504, \dodoi{10.1088/1538-3873/aaaaa8}

\bibitem[{Garhart {et~al.}(2020)Garhart, Deming, Mandell, Knutson, Wallack, Burrows, Fortney, Hood, Seay, Sing, Benneke, Fraine, Kataria, Lewis, Madhusudhan, McCullough, Stevenson, \& Wakeford}]{garhart_statistical_2020}
Garhart, E., Deming, D., Mandell, A., {et~al.} 2020, {\textbackslash}aj, 159, 137, \dodoi{10.3847/1538-3881/ab6cff}

\bibitem[{Harre {et~al.}(2023)Harre, Smith, Hirano, Csizmadia, Triaud, \& Anderson}]{harre_orbit_2023}
Harre, J.-V., Smith, A. M.~S., Hirano, T., {et~al.} 2023, The Astronomical Journal, 166, 159, \dodoi{10.3847/1538-3881/acf46d}

\bibitem[{Hellier {et~al.}(2017)Hellier, Anderson, Collier~Cameron, Delrez, Gillon, Jehin, Lendl, Maxted, Neveu-VanMalle, Pepe, Pollacco, Queloz, Ségransan, Smalley, Southworth, Triaud, Udry, Wagg, \& West}]{hellier_wasp-south_2017}
Hellier, C., Anderson, D.~R., Collier~Cameron, A., {et~al.} 2017, Monthly Notices of the Royal Astronomical Society, 465, 3693, \dodoi{10.1093/mnras/stw3005}

\bibitem[{Hinz {et~al.}(1998)Hinz, Angel, Hoffmann, McCarthy, McGuire, Cheselka, Hora, \& Woolf}]{hinz_imaging_1998}
Hinz, P.~M., Angel, J. R.~P., Hoffmann, W.~F., {et~al.} 1998, Nature, 395, 251, \dodoi{10.1038/26172}

\bibitem[{Hoeijmakers {et~al.}(2020)Hoeijmakers, Cabot, Zhao, Buchhave, Tronsgaard, Kitzmann, Grimm, Cegla, Bourrier, Ehrenreich, Heng, Lovis, \& Fischer}]{hoeijmakers_high-resolution_2020}
Hoeijmakers, H.~J., Cabot, S. H.~C., Zhao, L., {et~al.} 2020, A\&A, 641, A120, \dodoi{10.1051/0004-6361/202037437}

\bibitem[{Hoeijmakers {et~al.}(2024)Hoeijmakers, Prinoth, Borsato, Thorsbro, Morris, jseideleso, \& TrubbleMods}]{bibiana_prinoth_2024_11506199}
Hoeijmakers, J., Prinoth, B., Borsato, N.~W., {et~al.} 2024, tayph, v0.1,  Zenodo, \dodoi{10.5281/zenodo.11506199}

\bibitem[{Hut(1981)}]{hut_tidal_1981}
Hut, P. 1981, Astronomy and Astrophysics, 99, 126.
\newblock \url{https://ui.adsabs.harvard.edu/abs/1981A&A....99..126H}

\bibitem[{Jordán {et~al.}(2020)Jordán, Brahm, Espinoza, Henning, Jones, Kossakowski, Sarkis, Trifonov, Rojas, Torres, Drass, Nandakumar, Barbieri, Davis, Wang, Bayliss, Bouma, Dragomir, Eastman, Daylan, Guerrero, Barclay, Ting, Henze, Ricker, Vanderspek, Latham, Seager, Winn, Jenkins, Wittenmyer, Bowler, Crossfield, Horner, Kane, Kielkopf, Morton, Plavchan, Tinney, Addison, Mengel, Okumura, Shahaf, Mazeh, Rabus, Shporer, Ziegler, Mann, \& Hart}]{jordan_toi_677_2020}
Jordán, A., Brahm, R., Espinoza, N., {et~al.} 2020, The Astronomical Journal, 159, 145, \dodoi{10.3847/1538-3881/ab6f67}

\bibitem[{Kausch {et~al.}(2015)Kausch, Noll, Smette, Kimeswenger, Barden, Szyszka, Jones, Sana, Horst, \& Kerber}]{kausch_molecfit_2015}
Kausch, W., Noll, S., Smette, A., {et~al.} 2015, A\&A, 576, A78.
\newblock \url{https://www.aanda.org/articles/aa/abs/2015/04/aa23909-14/aa23909-14.html}

\bibitem[{Kitzmann {et~al.}(2023)Kitzmann, Stock, \& Patzer}]{kitzmann_fastchem_2023}
Kitzmann, D., Stock, J.~W., \& Patzer, A. B.~C. 2023, Monthly Notices of the Royal Astronomical Society, \dodoi{10.1093/mnras/stad3515}

\bibitem[{Lainey {et~al.}(2017)Lainey, Jacobson, Tajeddine, Cooper, Murray, Robert, Tobie, Guillot, Mathis, Remus, Desmars, Arlot, De~Cuyper, Dehant, Pascu, Thuillot, Le~Poncin-Lafitte, \& Zahn}]{lainey_new_2017}
Lainey, V., Jacobson, R.~A., Tajeddine, R., {et~al.} 2017, Icarus, 281, 286, \dodoi{10.1016/j.icarus.2016.07.014}

\bibitem[{Lee {et~al.}(2022)Lee, Prinoth, Kitzmann, Tsai, Hoeijmakers, Borsato, \& Heng}]{lee_mantis_2022}
Lee, E. K.~H., Prinoth, B., Kitzmann, D., {et~al.} 2022, MNRAS, 517, 240, \dodoi{10.1093/mnras/stac2246}

\bibitem[{Lodders(2010)}]{lodders_exoplanet_2010}
Lodders, K. 2010, in Formation and {Evolution} of {Exoplanets} (John Wiley \& Sons, Ltd), 157--186, \dodoi{10.1002/9783527629763.ch8}

\bibitem[{{Marconi} {et~al.}(2021){Marconi}, {Abreu}, {Adibekyan}, {Aliverti}, {Allende Prieto}, {Amado}, {Amate}, {Artigau}, {Augusto}, {Barros}, {Becerril}, {Benneke}, {Bergin}, {Berio}, {Bezawada}, {Boisse}, {Bonfils}, {Bouchy}, {Broeg}, {Cabral}, {Calvo-Ortega}, {Canto Martins}, {Chazelas}, {Chiavassa}, {Christensen}, {Cirami}, {Coretti}, {Covino}, {Cresci}, {Cristiani}, {Cunha Parro}, {Cupani}, {de Castro Le{\~a}o}, {Renan de Medeiros}, {Furlande Souza}, {Di Marcantonio}, {Di Varano}, {D'Odorico}, {Doyon}, {Drass}, {Figueira}, {Belen Fragoso}, {Uldall Fynbo}, {Gallo}, {Genoni}, {Gonz{\'a}lez Hern{\'a}ndez}, {Haehnelt}, {Hlavacek-Larrondo}, {Hughes}, {Huke}, {Humphrey}, {Kjeldsen}, {Korn}, {Kouach}, {Landoni}, {Liske}, {Lovis}, {Lunney}, {Maiolino}, {Malo}, {Marquart}, {Martins}, {Mason}, {Molaro}, {Monnier}, {Monteiro}, {Mordasini}, {Morris}, {Mucciarelli}, {Murray}, {Niedzielski}, {Nunes}, {Oliva}, {Origlia}, {Pall{\'e}}, {Pariani}, {Parr-Burman}, {Pe{\~n}ate}, {Pepe}, {Pinna}, {Piskunov}, {Rasilla
  Pi{\~n}eiro}, {Rebolo}, {Rees}, {Reiners}, {Riva}, {Romano}, {Rousseau}, {Sanna}, {Santos}, {Sarajlic}, {Shen}, {Sortino}, {Sosnowska}, {Sousa}, {Stempels}, {Strassmeier}, {Tenegi}, {Tozzi}, {Udry}, {Valenziano}, {Vanzi}, {Weber}, {Woche}, {Xompero}, {Zackrisson}, \& {Zapatero Osorio}}]{Marconi2021}
{Marconi}, A., {Abreu}, M., {Adibekyan}, V., {et~al.} 2021, The Messenger, 182, 27, \dodoi{10.18727/0722-6691/5219}

\bibitem[{Marley {et~al.}(2013)Marley, Ackerman, Cuzzi, \& Kitzmann}]{marley_clouds_2013}
Marley, M.~S., Ackerman, A.~S., Cuzzi, J.~N., \& Kitzmann, D. 2013, Clouds and {Hazes} in {Exoplanet} {Atmospheres}, \dodoi{10.2458/azu_uapress_9780816530595-ch015}

\bibitem[{McLaughlin(1924)}]{mclaughlin_results_1924}
McLaughlin, D.~B. 1924, The Astrophysical Journal, 60, 22, \dodoi{10.1086/142826}

\bibitem[{Mollière {et~al.}(2019)Mollière, Wardenier, van Boekel, Henning, Molaverdikhani, \& Snellen}]{molliere_petitradtrans_2019}
Mollière, P., Wardenier, J.~P., van Boekel, R., {et~al.} 2019, Astronomy \& Astrophysics, 627, A67, \dodoi{10.1051/0004-6361/201935470}

\bibitem[{Morley {et~al.}(2013)Morley, Fortney, Kempton, Marley, Vissher, \& Zahnle}]{morley_quantitatively_2013}
Morley, C.~V., Fortney, J.~J., Kempton, E. M.-R., {et~al.} 2013, The Astrophysical Journal, 775, 33, \dodoi{10.1088/0004-637X/775/1/33}

\bibitem[{Moses(2014)}]{moses_chemical_2014}
Moses, J.~I. 2014, Philosophical Transactions of the Royal Society of London Series A, 372, 20130073, \dodoi{10.1098/rsta.2013.0073}

\bibitem[{{Orell-Miquel} {et~al.}(2022){Orell-Miquel}, {Murgas}, {Pall{\'e}}, {Lamp{\'o}n}, {L{\'o}pez-Puertas}, {Sanz-Forcada}, {Nagel}, {Kaminski}, {Casasayas-Barris}, {Nortmann}, {Luque}, {Molaverdikhani}, {Sedaghati}, {Caballero}, {Amado}, {Bergond}, {Czesla}, {Hatzes}, {Henning}, {Khalafinejad}, {Montes}, {Morello}, {Quirrenbach}, {Reiners}, {Ribas}, {S{\'a}nchez-L{\'o}pez}, {Schweitzer}, {Stangret}, {Yan}, \& {Zapatero Osorio}}]{Orell-Miquel2022}
{Orell-Miquel}, J., {Murgas}, F., {Pall{\'e}}, E., {et~al.} 2022, \aap, 659, A55, \dodoi{10.1051/0004-6361/202142455}

\bibitem[{Palle {et~al.}(2023)Palle, Biazzo, Bolmont, Molliere, Poppenhaeger, Birkby, Brogi, Chauvin, Chiavassa, Hoeijmakers, Lellouch, Lovis, Maiolino, Nortmann, Parviainen, Pino, Turbet, Wender, Albrecht, Antoniucci, Barros, Beaudoin, Benneke, Boisse, Bonomo, Borsa, Brandeker, Brandner, Buchhave, Cheffot, Deborde, Debras, Doyon, Di~Marcantonio, Giacobbe, Gonzalez~Hernandez, Helled, Kreidberg, Machado, Maldonado, Marconi, Canto~Martins, Miceli, Mordasini, N'Diaye, Niedzielski, Nisini, Origlia, Peroux, Pietrow, Pinna, Rauscher, Reffert, Rousselot, Sanna, Simonnin, Suarez~Mascareno, Zanutta, \& Zechmeister}]{palle_ground-breaking_2023}
Palle, E., Biazzo, K., Bolmont, E., {et~al.} 2023, Ground-breaking {Exoplanet} {Science} with the {ANDES} spectrograph at the {ELT}, \dodoi{10.48550/arXiv.2311.17075}

\bibitem[{Pelletier {et~al.}(2021)Pelletier, Benneke, Darveau-Bernier, Boucher, Cook, Piaulet, Coulombe, Artigau, Lafrenière, Delisle, Allart, Doyon, Donati, Fouqué, Moutou, Cadieux, Delfosse, Hébrard, Martins, Martioli, \& Vandal}]{pelletier_where_2021}
Pelletier, S., Benneke, B., Darveau-Bernier, A., {et~al.} 2021, The Astronomical Journal, 162, 73, \dodoi{10.3847/1538-3881/ac0428}

\bibitem[{Pepe {et~al.}(2021)Pepe, Cristiani, Rebolo, Santos, Dekker, Cabral, Di~Marcantonio, Figueira, Curto, Lovis, Mayor, Mégevand, Molaro, Riva, Osorio, Amate, Manescau, Pasquini, Zerbi, Adibekyan, Abreu, Affolter, Alibert, Aliverti, Allart, Prieto, Álvarez, Alves, Avila, Baldini, Bandy, Barros, Benz, Bianco, Borsa, Bourrier, Bouchy, Broeg, Calderone, Cirami, Coelho, Conconi, Coretti, Cumani, Cupani, D'Odorico, Damasso, Deiries, Delabre, Demangeon, Dumusque, Ehrenreich, Faria, Fragoso, Genolet, Genoni, Santos, Hernández, Hughes, Iwert, Kerber, Knudstrup, Landoni, Lavie, Lillo-Box, Lizon, Maire, Martins, Mehner, Micela, Modigliani, Monteiro, Monteiro, Moschetti, Murphy, Nunes, Oggioni, Oliveira, Oshagh, Pallé, Pariani, Poretti, Rasilla, Rebordão, Redaelli, Tschudi, Santin, Santos, Ségransan, Schmidt, Segovia, Sosnowska, Sozzetti, Sousa, Spanò, Mascareño, Tabernero, Tenegi, Udry, \& Zanutta}]{pepe_espressovlt_2021}
Pepe, F., Cristiani, S., Rebolo, R., {et~al.} 2021, A\&A, 645, A96, \dodoi{10.1051/0004-6361/202038306}

\bibitem[{{Petrovich}(2015)}]{Petrovich2015}
{Petrovich}, C. 2015, \apj, 805, 75, \dodoi{10.1088/0004-637X/805/1/75}

\bibitem[{Pino {et~al.}(2018)Pino, Ehrenreich, Allart, Lovis, Brogi, Malik, Nascimbeni, Pepe, \& Piotto}]{pino_diagnosing_2018}
Pino, L., Ehrenreich, D., Allart, R., {et~al.} 2018, Astronomy and Astrophysics, 619, A3, \dodoi{10.1051/0004-6361/201832986}

\bibitem[{Prinoth(2024{\natexlab{a}})}]{bibiana_prinoth_RV}
Prinoth, B. 2024{\natexlab{a}}, Radial Velocity Trace Estimator, v1,  Zenodo, \dodoi{10.5281/zenodo.11505470}

\bibitem[{Prinoth(2024{\natexlab{b}})}]{bibiana_prinoth_ExoSim}
---. 2024{\natexlab{b}}, Exo Atmo Sim, v1,  Zenodo, \dodoi{10.5281/zenodo.11505486}

\bibitem[{Prinoth {et~al.}(2022)Prinoth, Hoeijmakers, Kitzmann, Sandvik, Seidel, Lendl, Borsato, Thorsbro, Anderson, Barrado, Kravchenko, Allart, Bourrier, Cegla, Ehrenreich, Fisher, Lovis, Guzmán-Mesa, Grimm, Hooton, Morris, Oreshenko, Pino, \& Heng}]{prinoth_titanium_2022}
Prinoth, B., Hoeijmakers, H.~J., Kitzmann, D., {et~al.} 2022, Nat Astron, 6, 449, \dodoi{10.1038/s41550-021-01581-z}

\bibitem[{Prinoth {et~al.}(2023)Prinoth, Hoeijmakers, Pelletier, Kitzmann, Morris, Seifahrt, Kasper, Korhonen, Burheim, Bean, Benneke, Borsato, Brady, Grimm, Luque, Stürmer, \& Thorsbro}]{prinoth_time-resolved_2023}
Prinoth, B., Hoeijmakers, H.~J., Pelletier, S., {et~al.} 2023, Astronomy and Astrophysics, 678, A182, \dodoi{10.1051/0004-6361/202347262}

\bibitem[{Prinoth {et~al.}(2024)Prinoth, Hoeijmakers, Morris, Lam, Kitzmann, Sedaghati, Seidel, Lee, Thorsbro, Borsato, Damasceno, Pelletier, \& Seifahrt}]{prinoth_atlas_2024}
Prinoth, B., Hoeijmakers, H.~J., Morris, B.~M., {et~al.} 2024, An atlas of resolved spectral features in the transmission spectrum of {WASP}-189 b with {MAROON}-{X}, \dodoi{10.48550/arXiv.2403.08863}

\bibitem[{Rice {et~al.}(2022)Rice, Wang, Wang, Stefánsson, Isaacson, Howard, Logsdon, Schweiker, Dai, Brinkman, Giacalone, \& Holcomb}]{rice_tendency_2022}
Rice, M., Wang, S., Wang, X.-Y., {et~al.} 2022, The Astronomical Journal, 164, 104, \dodoi{10.3847/1538-3881/ac8153}

\bibitem[{Rossiter(1924)}]{rossiter_detection_1924}
Rossiter, R.~A. 1924, The Astrophysical Journal, 60, 15, \dodoi{10.1086/142825}

\bibitem[{Rothman {et~al.}(2013)Rothman, Gordon, Babikov, Barbe, Chris~Benner, Bernath, Birk, Bizzocchi, Boudon, Brown, Campargue, Chance, Cohen, Coudert, Devi, Drouin, Fayt, Flaud, Gamache, Harrison, Hartmann, Hill, Hodges, Jacquemart, Jolly, Lamouroux, Le~Roy, Li, Long, Lyulin, Mackie, Massie, Mikhailenko, Müller, Naumenko, Nikitin, Orphal, Perevalov, Perrin, Polovtseva, Richard, Smith, Starikova, Sung, Tashkun, Tennyson, Toon, Tyuterev, \& Wagner}]{rothman_hitran2012_2013}
Rothman, L.~S., Gordon, I.~E., Babikov, Y., {et~al.} 2013, Journal of Quantitative Spectroscopy and Radiative Transfer, 130, 4, \dodoi{10.1016/j.jqsrt.2013.07.002}

\bibitem[{Rustamkulov {et~al.}(2022)Rustamkulov, Sing, Mukherjee, May, Kirk, Schlawin, Line, Piaulet, Carter, Batalha, Goyal, López-Morales, Lothringer, MacDonald, Moran, Stevenson, Wakeford, Espinoza, Bean, Batalha, Benneke, Berta-Thompson, Crossfield, Gao, Kreidberg, Powell, Cubillos, Gibson, Leconte, Molaverdikhani, Nikolov, Parmentier, Roy, Taylor, Turner, Wheatley, Aggarwal, Ahrer, Alam, Alderson, Allen, Banerjee, Barat, Barrado, Barstow, Bell, Blecic, Brande, Casewell, Changeat, Chubb, Crouzet, Daylan, Decin, Désert, Mikal-Evans, Feinstein, Flagg, Fortney, Harrington, Heng, Hong, Hu, Iro, Kataria, Kempton, Krick, Lendl, Lillo-Box, Louca, Lustig-Yaeger, Mancini, Mansfield, Mayne, Miguel, Morello, Ohno, Palle, de~la Roche, Rackham, Radica, Ramos-Rosado, Redfield, Rogers, Shkolnik, Southworth, Teske, Tremblin, Tucker, Venot, Waalkes, Welbanks, Zhang, \& Zieba}]{rustamkulov_early_2022}
Rustamkulov, Z., Sing, D.~K., Mukherjee, S., {et~al.} 2022, Early {Release} {Science} of the exoplanet {WASP}-39b with {JWST} {NIRSpec} {PRISM},  arXiv, \dodoi{10.48550/arXiv.2211.10487}

\bibitem[{Sedaghati {et~al.}(2023)Sedaghati, Jordán, Brahm, Muñoz, Petrovich, \& Hobson}]{sedaghati_orbital_2023}
Sedaghati, E., Jordán, A., Brahm, R., {et~al.} 2023, The Astronomical Journal, 166, 130, \dodoi{10.3847/1538-3881/acea84}

\bibitem[{Sha {et~al.}(2021)Sha, Huang, Shporer, Rodriguez, Vanderburg, Brahm, Hagelberg, Matthews, Ziegler, Livingston, Stassun, Wright, Crane, Espinoza, Bouchy, Bakos, Collins, Zhou, Bieryla, Hartman, Wittenmyer, Nielsen, Plavchan, Bayliss, Sarkis, Tan, Cloutier, Mancini, Jordán, Wang, Henning, Narita, Penev, Teske, Kane, Mann, Addison, Tamura, Horner, Barbieri, Burt, Díaz, Crossfield, Dragomir, Drass, Feinstein, Zhang, Hart, Kielkopf, Jensen, Montet, Ottoni, Schwarz, Rojas, Nespral, Torres, Mengel, Udry, Zapata, Snoddy, Okumura, Ricker, Vanderspek, Latham, Winn, Seager, Jenkins, Colón, Henze, Krishnamurthy, Ting, Vezie, \& Villanueva}]{sha_toi-954_2021}
Sha, L., Huang, C.~X., Shporer, A., {et~al.} 2021, The Astronomical Journal, 161, 82, \dodoi{10.3847/1538-3881/abd187}

\bibitem[{Smette {et~al.}(2015)Smette, Sana, Noll, Horst, Kausch, Kimeswenger, Barden, Szyszka, Jones, Gallenne, \& {others}}]{smette_molecfit_2015}
Smette, A., Sana, H., Noll, S., {et~al.} 2015, A\&A, 576, A77

\bibitem[{Smith {et~al.}(2014)Smith, Anderson, Armstrong, Barros, Bonomo, Bouchy, Brown, Cameron, Delrez, Faedi, Gillon, Chew, Hébrard, Jehin, Lendl, Louden, Maxted, Montagnier, Neveu-VanMalle, Osborn, Pepe, Pollacco, Queloz, Rostron, Segransan, Smalley, Triaud, Turner, Udry, Walker, West, \& Wheatley}]{smith_wasp-104b_2014}
Smith, A. M.~S., Anderson, D.~R., Armstrong, D.~J., {et~al.} 2014, Astronomy \& Astrophysics, 570, A64, \dodoi{10.1051/0004-6361/201424752}

\bibitem[{Snellen {et~al.}(2010)Snellen, de~Kok, de~Mooij, \& Albrecht}]{snellen_orbital_2010}
Snellen, I. A.~G., de~Kok, R.~J., de~Mooij, E. J.~W., \& Albrecht, S. 2010, Nature, 465, 1049, \dodoi{10.1038/nature09111}

\bibitem[{Stassun {et~al.}(2017)Stassun, Collins, \& Gaudi}]{stassun_accurate_2017}
Stassun, K.~G., Collins, K.~A., \& Gaudi, B.~S. 2017, The Astronomical Journal, 153, 136, \dodoi{10.3847/1538-3881/aa5df3}

\bibitem[{Stock {et~al.}(2022)Stock, Kitzmann, \& Patzer}]{stock_fastchem_2022}
Stock, J.~W., Kitzmann, D., \& Patzer, A. B.~C. 2022, MNRAS, 517, 4070, \dodoi{10.1093/mnras/stac2623}

\bibitem[{Stock {et~al.}(2018)Stock, Kitzmann, Patzer, \& Sedlmayr}]{stock_fastchem_2018}
Stock, J.~W., Kitzmann, D., Patzer, A. B.~C., \& Sedlmayr, E. 2018, MNRAS, 479, 865

\bibitem[{Tsai {et~al.}(2023)Tsai, Parmentier, Mendonça, Tan, Deitrick, Hammond, Savel, Zhang, Pierrehumbert, \& Schwieterman}]{tsai_global_2023}
Tsai, S.-M., Parmentier, V., Mendonça, J.~M., {et~al.} 2023, Global {Chemical} {Transport} on {Hot} {Jupiters}: {Insights} from {2D} {VULCAN} photochemical model,  arXiv, \dodoi{10.48550/arXiv.2310.17751}

\bibitem[{Visscher(2012)}]{visscher_chemical_2012}
Visscher, C. 2012, The Astrophysical Journal, 757, 5, \dodoi{10.1088/0004-637X/757/1/5}

\bibitem[{Visscher \& Moses(2011)}]{visscher_quenching_2011}
Visscher, C., \& Moses, J.~I. 2011, The Astrophysical Journal, 738, 72, \dodoi{10.1088/0004-637X/738/1/72}

\bibitem[{Wildi {et~al.}(2017)Wildi, Blind, Reshetov, Hernandez, Genolet, Conod, Sordet, Segovilla, Rasilla, Brousseau, Thibault, Delabre, Bandy, Sarajlic, Cabral, Bovay, Vallée, Bouchy, Doyon, Artigau, Pepe, Hagelberg, Melo, Delfosse, Figueira, Santos, González~Hernández, de~Medeiros, Rebolo, Broeg, Benz, Boisse, Malo, Käufl, \& Saddlemyer}]{wildi_nirps_2017}
Wildi, F., Blind, N., Reshetov, V., {et~al.} 2017, 10400, 1040018, \dodoi{10.1117/12.2275660}

\bibitem[{Winn(2010)}]{winn_transits_2010}
Winn, J.~N. 2010, arXiv e-prints, arXiv:1001.2010

\bibitem[{Wright {et~al.}(2023)Wright, Rice, Wang, Hixenbaugh, \& Wang}]{wright_soles_2023}
Wright, J., Rice, M., Wang, X.-Y., Hixenbaugh, K., \& Wang, S. 2023, The Astronomical Journal, 166, 217, \dodoi{10.3847/1538-3881/ad0131}

\bibitem[{Yurchenko \& Tennyson(2014)}]{yurchenko_exomol_2014}
Yurchenko, S.~N., \& Tennyson, J. 2014, Monthly Notices of the Royal Astronomical Society, 440, 1649, \dodoi{10.1093/mnras/stu326}

\end{thebibliography}
\bibliographystyle{aasjournal}

\newpage
\appendix


\section{Cross-correlation templates}

\begin{figure}[h!]
    \centering
    \includegraphics[width=\textwidth]{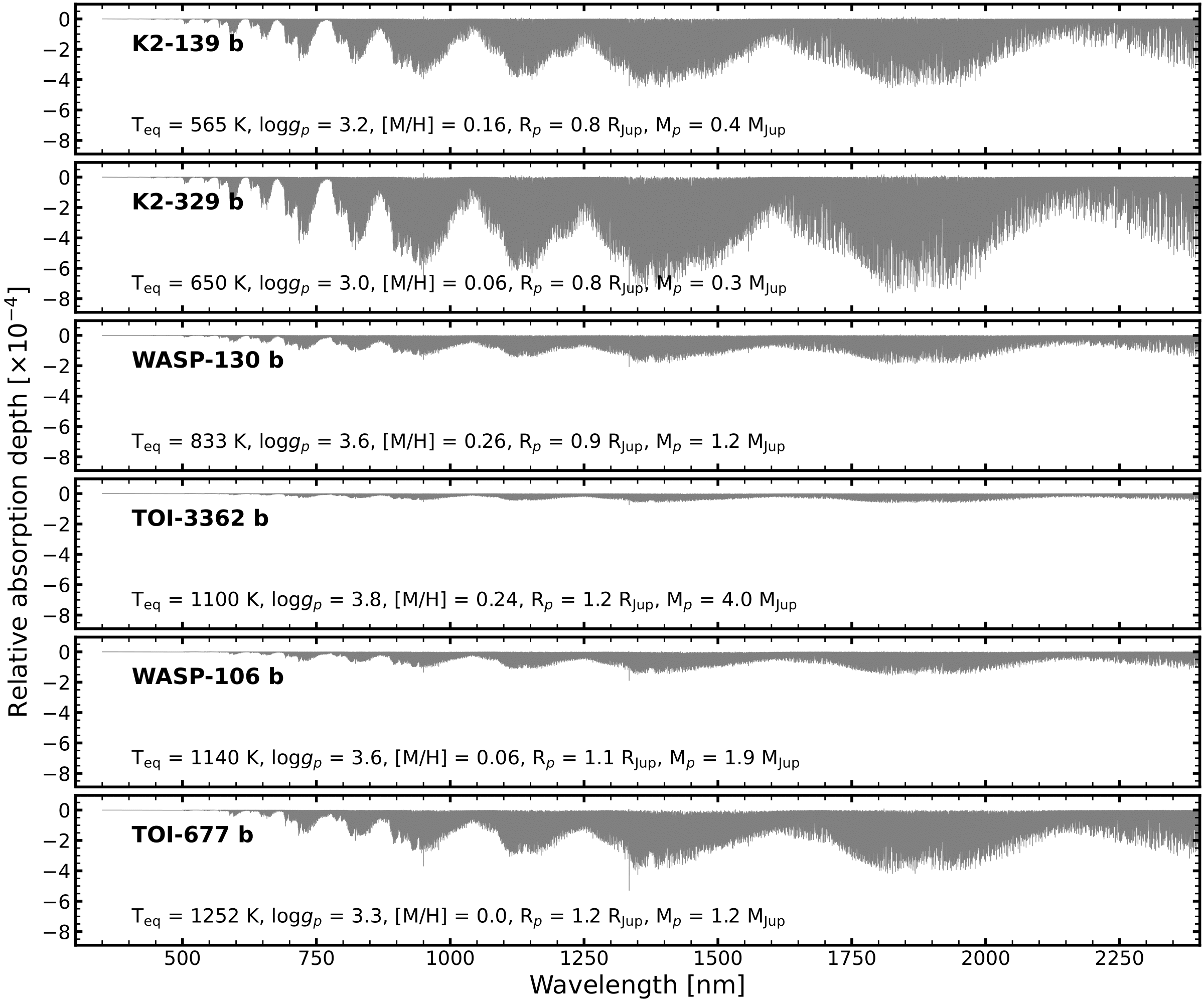}
    \caption{Cross-correlation templates for \water covering the wavelength range of ESPRESSO and ANDES for \allplanets. The cross-correlation templates assume the parameters indicated in the panels and were generated using \texttt{FastChem Cond} for the abundance profiles and \texttt{petitRADTRANS} for the computation. Apart from \water, the template includes continuum absorption by He and \ch{H2}.}
    \label{fig:cc}
\end{figure}
\newpage
\section{RV Trace Estimator}
\label{app:traces}

\begin{figure}
    \centering
    \includegraphics[width=\linewidth]{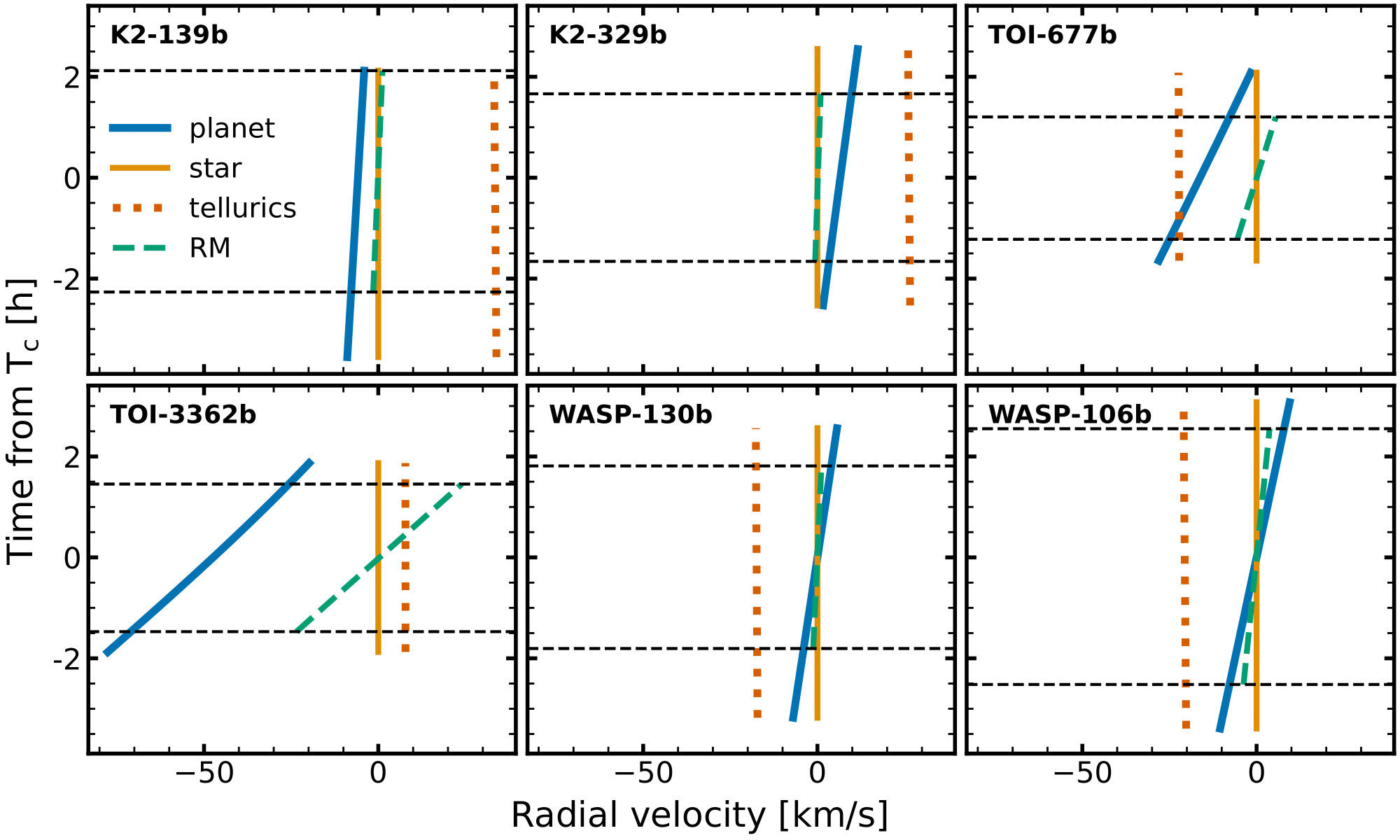}
    \caption{Radial velocities of the planet (blue, solid), star (orange, solid), RM effect (green, dashed) and telluric contamination (orange, dotted) for \allplanets in the rest frame of the system. The dashed black lines indicate the transit contact times.}
    \label{fig:RVTraceEstimator}
\end{figure}

Planning observations to optimise the velocity components of transiting planets is crucial for accessing the planetary signal while minimising contamination from the star, residual telluric contamination, and the RM effect. While the stellar component and RM effect remain stationary, the telluric contamination depends on the observing date, particularly the location of Earth in its orbit around the Sun (known as Barycentric Earth Radial Velocity or BERV). The code \texttt{RVTraceEstimator} allows us to estimate the observed radial velocities of the planet, star, RM effect, and telluric contamination, provided the system's configuration is known. Fig.\,\ref{fig:RVTraceEstimator} shows the predicted radial velocities of these components for the six planets in this study.

\newpage

\section{Parameters}

\begin{table}[h!]
\movetabledown=2.75in
\begin{rotatetable}
\caption{Planetary, stellar and orbital parameters.}
\begin{center}
\label{tab:planets}
\def\arraystretch{1.2}
{\small
\begin{tabular}{lllllll}
    \toprule
    Parameter                                                 &   K2-139\,b                            &  K2-329\,b                             & TOI-3362\,b                           & WASP-130\,b                               & WASP-106\,b                     & TOI-677\,b \\
    \toprule
    \multicolumn{7}{c}{Stellar parameters from ESPRESSO spectral synthesis with {\tt zaspe} \citep{Brahm2017}} \\ \midrule
    Star radius $R_\ast$ [$R_{\odot}$]                        & $0.85\pm0.01$                          & $0.81\pm 0.01$                         & $1.75 \pm 0.02$                       & $1.00 \pm 0.01$                           & $1.41\pm0.02$                   & $1.281 \pm 0.012$ \\ 
    Star mass $M_\ast$ [$M_{\odot}$]                          & $0.92\pm0.03$                          & \numpm{0.88}{+0.02}{-0.03}             & $1.53 \pm 0.02$                       & $1.05 \pm 0.03$                           & $1.22\pm0.03$                   & \numpm{1.158}{+0.029}{-0.027}\\ 
    Proj. rot. velocity $v\sin{I_\ast}$ [\si{\km\per\second}] & $3.6 \pm 0.3$                          & $2.9 \pm 0.3$                          & \numpm{20.2}{+1.7}{-1.6}              & $3.5 \pm 0.3$                             & $7.5 \pm 0.4$                   & $7.42 \pm 0.5$\\ 
    Systemic velocity $v_{\rm sys}$ [\si{\km\per\second}]     & $-31.3575 \pm 0.0064 $                 & $-19.25 \pm 2.88$                      & \numpm{7.637}{+0.035}{-0.033}         & $1.462 \pm 0.002$                         & $17.24744\pm0.00019$            & \numpm{37.94068}{+0.00497}{-0.00535}\\ 
    Surface gravity log$g_\ast$ [\si{\cm\per\s\squared}]      & $4.54 \pm 0.02$                        & $4.57 \pm 0.02$                        & $4.14 \pm 0.01$                       & $4.46 \pm 0.02$                           & $4.23 \pm 0.02$                 & \numpm{4.286}{+0.016}{-0.015}\\ 
    Effective temperature $T_{\rm eff}$ [$\si{\kelvin}$]      & $5360 \pm 70$                          & $5300 \pm 60$                          & $6800 \pm 100$                        & $5680 \pm 80$                             & $6270 \pm 100$                  & $6295 \pm 80$\\ 
    Metallicity [Fe/H] [dex]                                  & $0.16 \pm 0.05$                        & $0.06 \pm 0.04$                        & $0.24 \pm 0.05$                       & $0.26 \pm 0.05$                           & $0.06 \pm 0.05$                 & $-0.02 \pm 0.05$\\ 
    Spectral type                                             & K0 V                                   & G                                      & G5                                    & G6                                        & F9 [9]                          & F \\
    V-band magnitude $m_V$ [mag]                              & 11.7                                   & 12.7                                   & 10.9                                  & 11.1                                      & 11.2                            & 9.82\\
    \midrule
    \multicolumn{6}{c}{Orbital, planetary and stellar parameters from modelling the RM effect} & \citet{sedaghati_orbital_2023} \\ \midrule
    Transit centre time $T_0$ {\tiny [BJD - 2450000]}         & \numpm{9766.73165}{+0.00098}{-0.00104}  & \numpm{9815.79516}{+0.00178}{-0.00177}  & \numpm{9940.76670}{+0.00101}{-0.00098} &\numpm{9739.58215}{+0.00117}{-0.00123}  & \numpm{10021.714040}{+0.001124}{-0.001112}&\numpm{9558.76931}{0.00070}{0.00066}\\ 
    Orbital semi-major axis $a$ [AU]                          & $0.169\pm0.005$                        & $0.101\pm0.004$                        & $0.160\pm0.010$                       & $0.081 \pm 0.005$                         & $0.094\pm0.003$                 & \numpm{0.0945}{+0.0095}{-0.0079}\\ 
    Scaled semi-major axis $a/R_\ast$                         & \numpm{42.92}{+0.87}{-0.86}            & \numpm{26.83}{+0.69}{-0.67}            & \numpm{19.60}{+1.03}{-0.98}           & $17.45 \pm 0.77$                          & \numpm{14.41}{+0.19}{-0.20}     & \numpm{15.86}{+1.58}{-1.32}\\ 
    Orbital inclination $i$ [$\deg$]                          & $89.64\pm0.01$                         & \numpm{89.20}{+0.28}{-0.21}            & \numpm{85.59}{+1.64}{-1.05}           & \numpm{87.38}{+0.26}{-0.24}               & \numpm{89.19}{+0.28}{-0.25}     & \numpm{84.80}{+0.80}{-0.79}\\ 
    Projected orbital obliquity $\lambda$ [$\deg$]            & $-17\pm7$                              & \numpm{24}{+13}{-9}                    & \numpm{2}{+4}{-5}                     & $-2\pm2$                                  & \numpm{-3}{+9}{-10}             & $0 \pm 1$\\ 
    Eclipse duration $T_{14}$ [h]                             & $4.72\pm0.09$                          & $3.58\pm0.16$                          & $2.85\pm0.07$                         & $3.80\pm0.18$                             & $5.23 \pm 0.08$                  & $2.52\pm0.07$\\ 
    Eccentricity $\epsilon$                                   & $0.16\pm0.03$                          & $0.08\pm0.04$                          & $0.72 \pm 0.01$                       & 0 (fixed)                                 & 0 (fixed)                       & $0.44 \pm 0.02$\\
    Argument of periastron $\omega$ [$\deg$]                  & $58\pm3$                               & $160\pm9$                              & $61 \pm 1$                            & 90 (fixed)                                & 90 (fixed)                      & $70 \pm 3$\\  
    Proj. rot. velocity $v\sin{I_\ast}$ [\si{\km\per\second}] & $1.39\pm0.06$                          & \numpm{0.89}{+0.09}{-0.08}             & \numpm{24.4}{+19.8}{-7.7}             & \numpm{1.65}{+0.28}{-0.30}                & \numpm{3.64}{+0.15}{-0.13}      & \numpm{6.91}{+1.32}{-1.20}\\ 
    Systemic velocity $v_{\rm sys}$ [\si{\km\per\second}]     & $-31.348\pm0.001$                      & $-17.141\pm0.001$                      & $7.637\pm0.004$                       & $1.3569\pm0.0003$                         & $17.135 \pm 0.001$              & $-37.941\pm0.005$\\ 
    \midrule
    \multicolumn{7}{c}{Orbital, planetary and stellar parameters from literature} \\ 
                                                              & \citet{barragan_k2-139_2018}           & \citet{sha_toi-954_2021}               & \citet{espinoza-retamal_aligned_2023} & \cite{hellier_wasp-south_2017}            & \citet{smith_wasp-104b_2014}    & \citet{jordan_toi_677_2020} \\        
    \midrule
    Period $P$ [d]                                            & $28.38236\pm0.00026$                   & $12.45512 \pm 0.00001$                 & $18.09537 \pm 0.00001$                & $11.55098 \pm 0.00001$                    & $9.28971\pm0.00001$             & $11.23660 \pm 0.00011$\\ 
    RV semi-amplitude $K$  [\si{\m\per\second}]               & \numpm{27.7}{+6.0}{-5.3}               & \numpm{24.6}{+1.6}{-1.8}               & $338 \pm 27$                          & $108 \pm 2$                               & $165 \pm 4$                     & $111.6 \pm 0.5$\\ 
    Radius ratio $R_p/R_\ast$                                 & \numpm{0.0961}{+0.0023}{-0.0015}       & $0.0968\pm0.0006$                      & $0.070 \pm 0.0001$                    & $0.096 \pm  0.012$                        & $0.0801\pm0.0045$               & $0.0942 \pm 0.0012$\\ 
    Planet radius $R_{\rm p}$ [$R_{\rm Jup}$]                 & \numpm{0.808}{+0.034}{-0.033}          & \numpm{0.774}{+0.026}{-0.024}          & $1.2 \pm 0.02$                        & $0.89 \pm 0.03$                           & \numpm{1.09}{+0.05}{-0.03}      & $1.174 \pm 0.018$\\ 
    Planet mass $M_{\rm p}$ [$M_{\rm Jup}$]                   & \numpm{0.387}{+0.083}{-0.075}          & \numpm{0.260}{+0.020}{-0.022}          & $4.0 \pm 0.4$                         & $1.23 \pm 0.04$                           & $1.93 \pm 0.08$                 & $1.24 \pm 0.07$ \\
    Eq. temperature $T_{\rm eq}$ [$\si{\kelvin}$]             & \numpm{565}{+48}{-32}                  & \numpm{650}{+53}{-70}                  & $800 - 2500$ [4]                      & $833 \pm 18$                              & $1140 \pm 29$                   & $1252 \pm 21$\\
    \bottomrule
\end{tabular}
}
\end{center}
\end{rotatetable}
\end{table}


\end{document}